\documentclass[12pt]{article}

\usepackage{dina4pcg}
\usepackage{epsfig}
\usepackage{amssymb}

\def\as{\alpha_S}
\def\z0{Z^0}
\def\mf2{\mu_F^2}
\def\mr2{\mu_R^2}
\def\q2{Q^2}
\def\etjet{E_T^{\rm jet}}
\def\asz{\as(\mz)}
\def\kt{k_T}
\def\g2{GeV$^2$}
\def\etjb{E^{\rm jet}_{T,{\rm B}}}
\def\etajb{\eta^{\rm jet}_{\rm B}}
\def\etarr{-1<\etajet<2}
\def\etajet{\eta^{\rm jet}}
\def\mjj{m_{\rm JJ}}
\def\oas{{\cal O}(\as)}
\def\oalphass{{\cal O}(\alpha\as)}
\def\oalphas2{{\cal O}(\alpha\as^2)}
\def\etaphi{\eta-\varphi}
\def\yc{y_{\rm cut}}
\def\ns{n_{\rm subjet}}
\def\asmz#1#2#3#4#5#6{\asz = #1\pm #2\ {\rm (stat)}\ ^{+#3}_{-#4}\ {\rm (exp)}\ ^{+#5}_{-#6}\ {\rm (th)}}
\def\Journal#1#2#3#4{{#1} {#2} (#3) #4}

\def\NPB{{\em Nucl. Phys.} {\bf B}}
\def\PLB{{\em Phys. Lett.}  {\bf B}}

\def\PRD{{\em Phys. Rev.} {\bf D}}

\def\ZPC{{\em Z. Phys.} {\bf C}}

\def\EPC{{\em Eur. Phys. Jour.} {\bf C}}

\def\CPC{{\em Comp. Phys. Comm.}}
\def\JPG{{\em J. Phys. {\bf G}}}
\def\etal{{\it et al}}
\def\mz{M_Z}
\def\colab#1{{#1 Collaboration}}
\def\conf#1#2#3#4{{paper #1}, {submitted to the #2}, {#3} {(#4)}}
\def\wp{W^{\pm}}

\begin{document}

\title{\bf Jet production in deep inelastic {\boldmath $ep$}
  scattering at HERA and determination of {\boldmath $\as$}}

\author{C. Glasman\\
University of Glasgow, UK}

\date{}

\maketitle

\begin{abstract}
Recent measurements of jet cross sections in neutral-current and
charged-current deep inelastic $ep$ scattering are presented. The results
of the QCD analyses on these measurements to determine the strong
coupling constant are also reported.
\end{abstract}

\section{Introduction}

Jet production in neutral-current (NC) and charged-current (CC) deep
inelastic $ep$ scattering (DIS) provides a test of perturbative QCD
(pQCD) calculations and of the electroweak sector of the Standard
Model (SM). Jet cross sections allow the determination of one of the
fundamental parameters of QCD, the strong coupling constant $\as$, and
help to constrain the parton densities in the proton. New particles or
interactions may be observed by deviations of the measured jet cross
sections with respect to the predictions.

\begin{figure}[h]
\setlength{\unitlength}{1.0cm}
\begin{picture} (18.0,5.5)
\put (0.0,0.5){\epsfig{figure=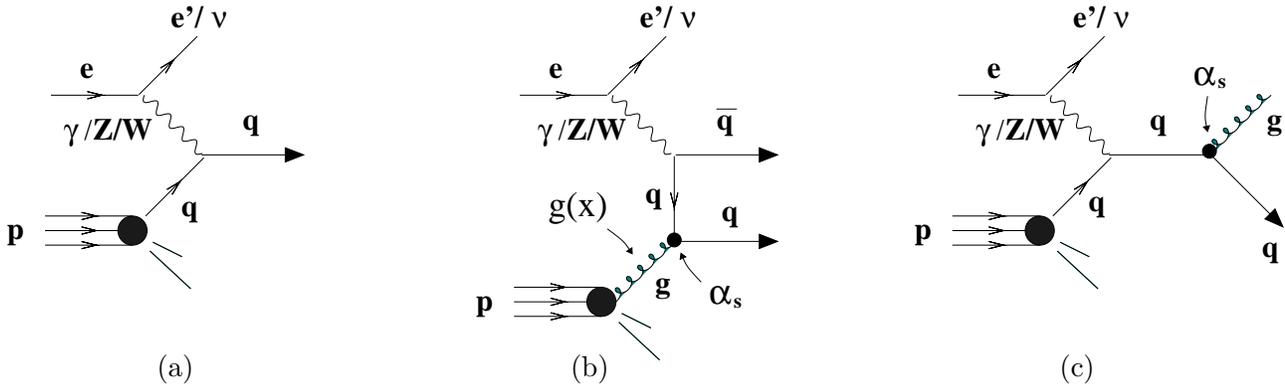,width=17cm}}
\put (2.0,0.3){\small (a)}
\put (7.5,0.3){\small (b)}
\put (14.0,0.3){\small (c)}
\end{picture}
\caption{\label{one}
{Quark-parton model (a), boson-gluon fusion (b) and QCD Compton (c)
diagrams.}}
\end{figure}

Up to leading order (LO) in $\as$, jet production in NC and CC DIS
proceeds via the quark-parton model (QPM) ($Vq\rightarrow q$, where
$V=\gamma$, $\z0$ or $\wp$), boson-gluon fusion (BGF) ($Vg\rightarrow
q\bar q$) and QCD-Compton (QCDC) ($Vq\rightarrow qg$) processes (see
figure~\ref{one}).

The jet production cross section is given in pQCD by the convolution
of the parton densities in the proton and the subprocess cross section,
$$d\sigma_{\rm jet}=\sum_{a=q,\bar q,g}\int dx\ f_a(x,\mf2)\ d\hat \sigma_a(x,\as(\mu_R),\mr2,\mf2),$$
where $x$ is the fraction of the proton momentum taken by the interacting
parton, $f_a$ are the parton distribution functions (PDFs) in the proton,
$\mu_F$ is the factorisation scale, $\hat\sigma_a$ is the subprocess
cross section and $\mu_R$ is the renormalisation scale.

The proton PDFs are determined from global fits to experimental
data (see e.g. \cite{cteq,mrs}) and parametrise the long-distance
structure of the target hadron. The subprocess cross section is
calculable in pQCD at any order and represents the short-distance
structure of the interaction.

The determination of $\as$ from measurements of cross sections in DIS
depends on the proton PDFs. Thus, to perform a QCD analysis, a functional
form of the PDFs must be assumed and the determination of $\as$ should be
made from observables with a small dependence on the PDFs, e.g. ratios of
observables. For precise measurements of $\as$, the experimental and
theoretical uncertainties should be small. Small experimental
uncertainties are obtained by measuring e.g. ratios of observables and
the theoretical uncertainties are reduced by including next-to-leading
order (NLO) QCD corrections to the cross section calculations.

\section{NLO QCD calculations}

Several programs are available to make NLO QCD calculations of jet
cross sections in DIS: DISENT \cite{disent}, MEPJET \cite{mepjet},
DISASTER++ \cite{disaster} and NLOJET \cite{nlojet}. The NLO corrections
include virtual corrections with internal particle loops and real
corrections with a third parton in the final state. The programs differ
in the treatment of the real corrections: DISENT, DISASTER++ and NLOJET
use the subtraction method whereas MEPJET uses the phase space slicing
method. DISENT and DISASTER++ are found to agree at the $2\%$ level, and
NLOJET is in good agreement with them. MEPJET agrees well with
the other programs at LO, but presents differences of the order of $5\%$
at NLO. MEPJET is the only program that includes $\z0$ and $W$ exchange
and only DISASTER++ allows the possibility to turn on the number of active
flavours as a function of the scale.

The calculation of jet cross sections at NLO in DIS depends on two scales,
$\mu_R$ and $\mu_F$. Two possible choices for these scales that have been
considered in the analyses presented here are $Q\equiv\sqrt{\q2}$,
where $\q2$ is the virtuality of the exchanged boson, and the
jet transverse energy, $\etjet$.

Since the NLO QCD calculations are for jets of partons and the
measurements are done at the hadron level, the calculations need to be
corrected for hadronisation effects. These effects have been estimated
using Monte Carlo models for parton radiation and hadronisation
\cite{cdm,lepto}.

The uncertainties of the QCD calculations include that due to terms beyond
NLO, which is usually estimated by varying the renormalisation scale by
factors between $1/2$ and $2$; it amounts to $5-10\%$ and translates
into $\sim 8\%$ in the determination of $\asz$. The uncertainties on
the value of $\asz$ and of the proton PDFs amount to $\sim 5\%$. The
size and uncertainty of the hadronisation corrections have been taken
into account: the hadronisation correction factor amounts to $\sim 5\%$
with an uncertainty of the order of $2\%$.

\section{Inclusive jet cross sections}

The use of inclusive jet cross sections in a QCD analysis presents several
advantages: inclusive jet cross sections are infrared insensitive and
better suited to test resummed calculations and the theoretical
uncertainties are smaller than for dijet cross sections.

\begin{figure}[h]
\setlength{\unitlength}{1.0cm}
\begin{picture} (18.0,15.0)
\put (-1.0,5.0){\epsfig{figure=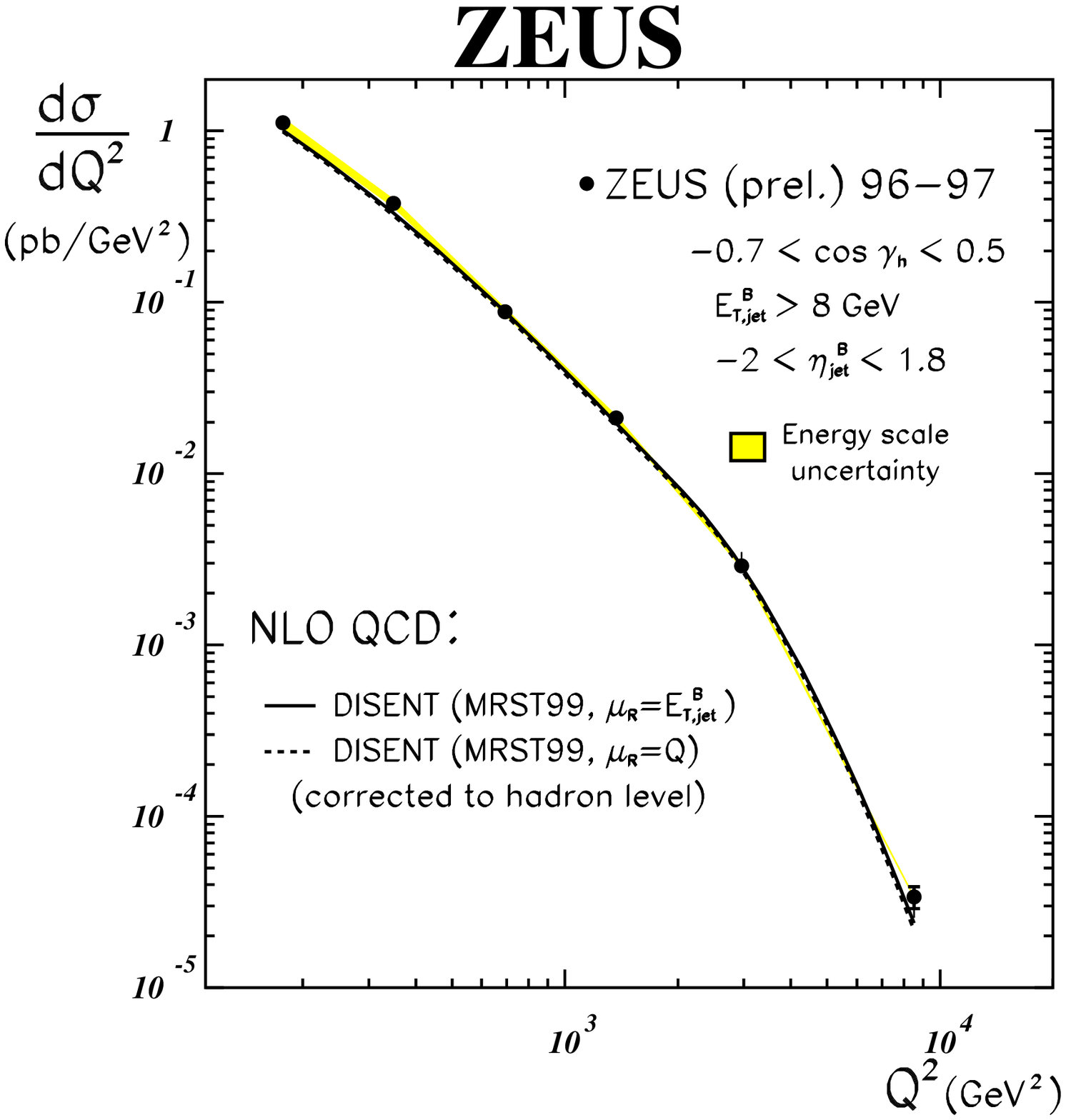,width=10cm}}
\put (-1.0,0.0){\epsfig{figure=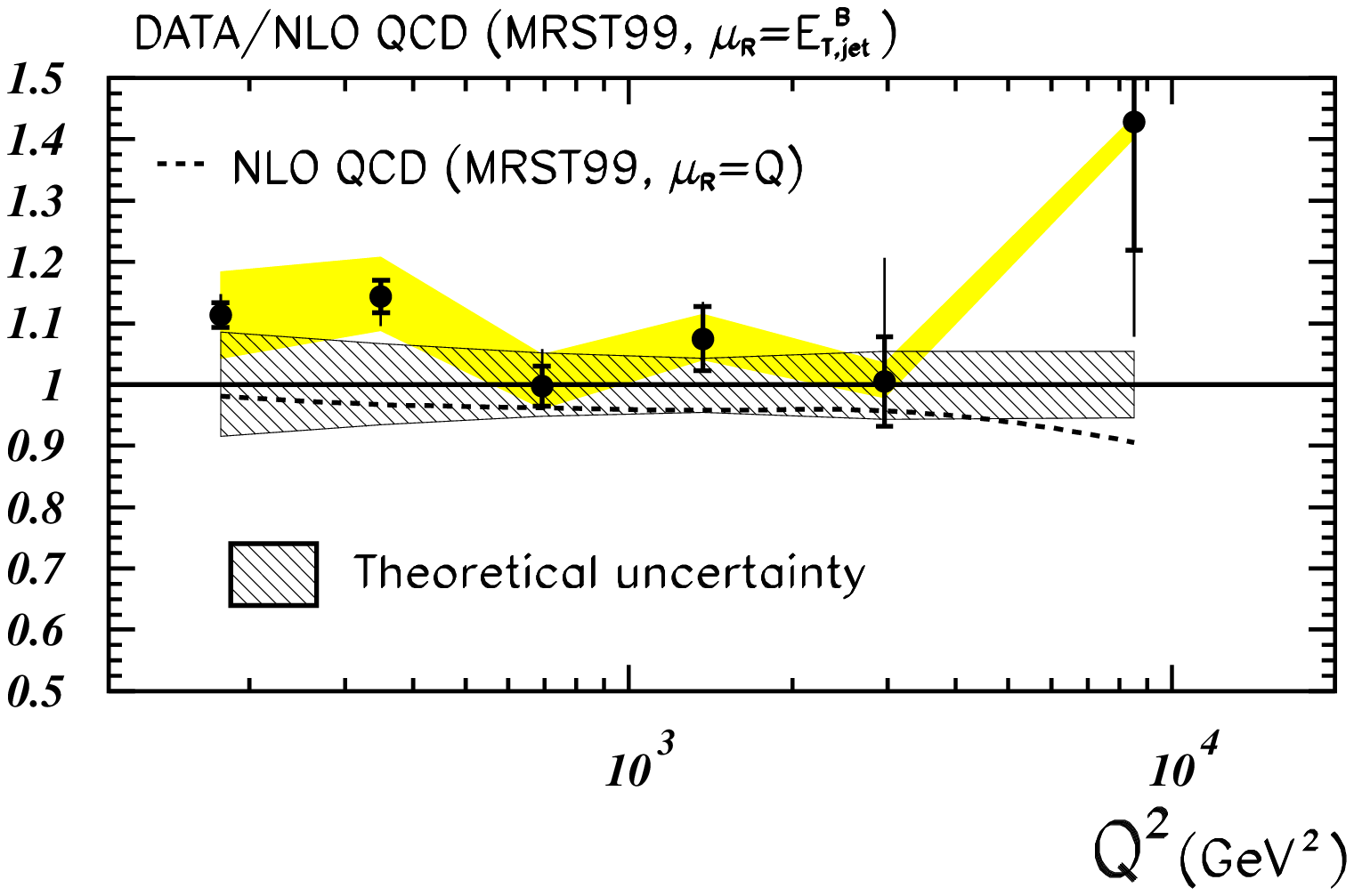,width=10cm}}
\put (8.0,5.0){\epsfig{figure=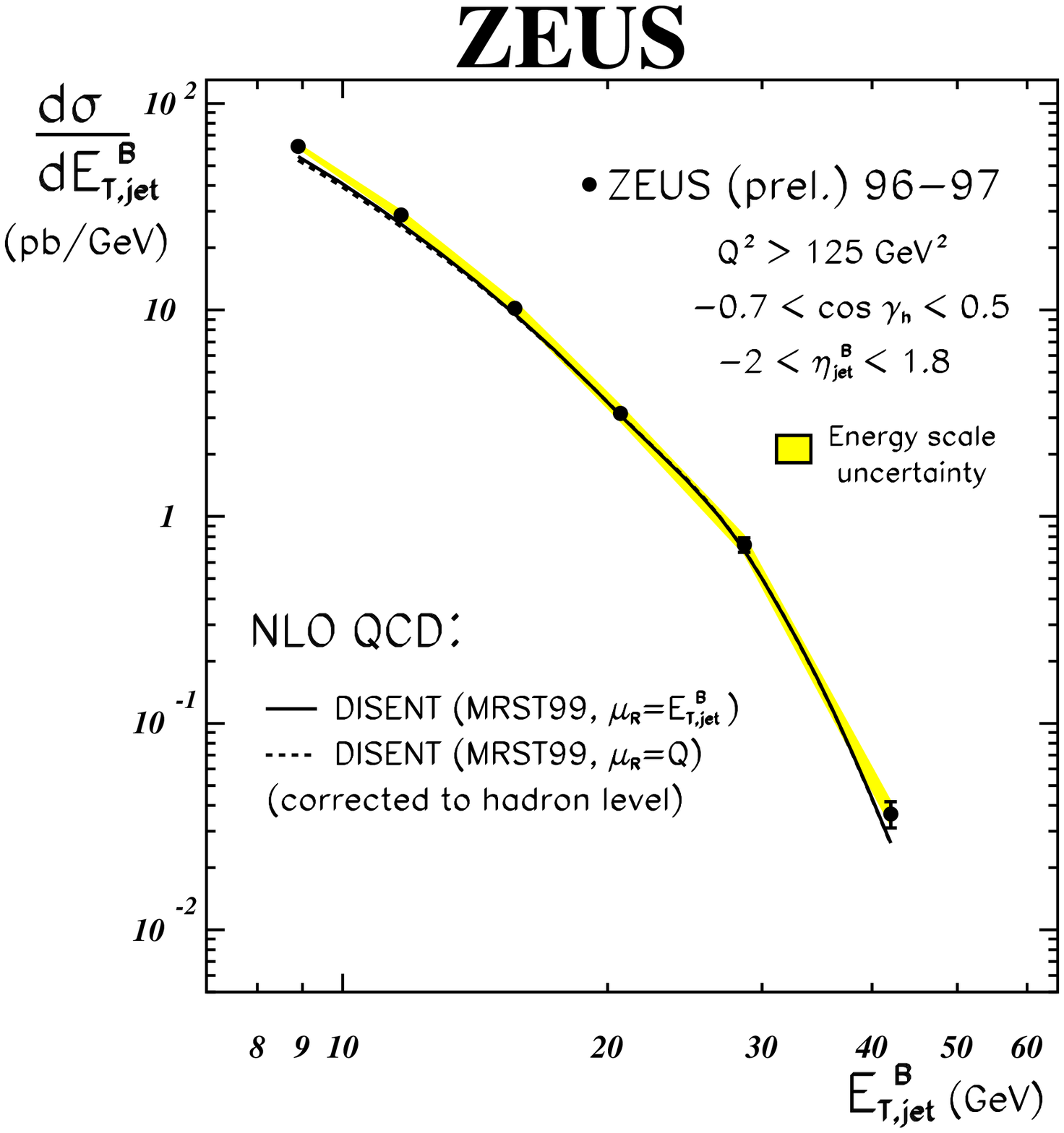,width=10cm}}
\put (8.0,0.0){\epsfig{figure=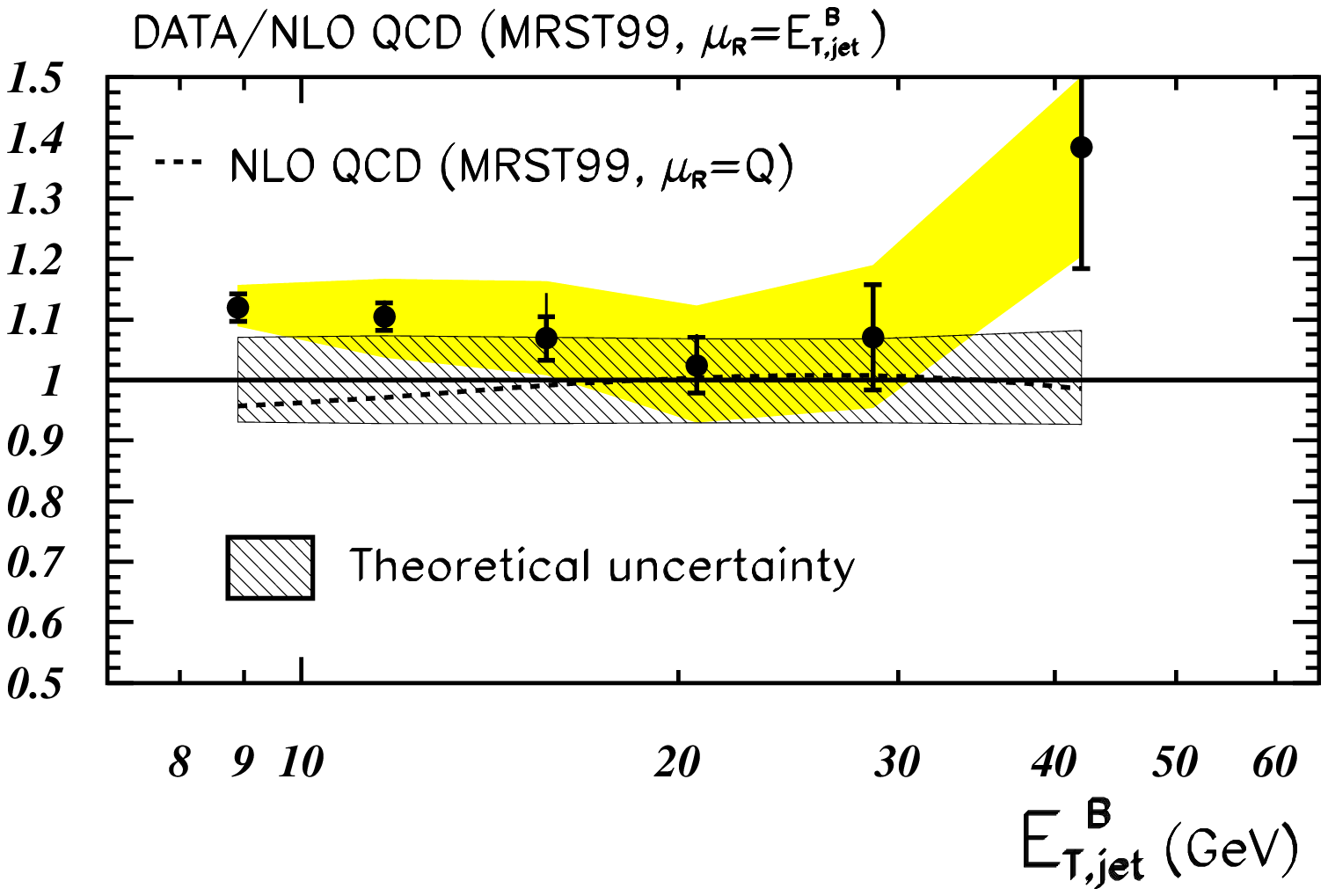,width=10cm}}
\put (7.0,13.5){\small (a)}
\put (7.0,4.3){\small (b)}
\put (16.0,13.5){\small (c)}
\put (16.0,4.3){\small (d)}
\end{picture}
\caption{\label{two}
{Inclusive jet cross sections in NC DIS as a function of $\q2$ (a) and
$\etjb$ (c). Ratio data/theory as a function of $\q2$ (b) and $\etjb$ 
(d).}}
\end{figure}

Inclusive jet cross sections in NC interactions have been measured
\cite{inclusive} in the Breit frame using the $\kt$ cluster algorithm
\cite{kt} in the longitudinally invariant inclusive mode for the kinematic
region of $\q2>125$ \g2, where $\q2$ is the virtuality of the exchanged
boson. These cross sections refer to jets of transverse energy
measured in the Breit frame, $\etjb$, above 8 GeV and jet
pseudorapidity also in the Breit frame in the range
$-2<\etajb<1.8$. The inclusive jet cross sections as a function of
$\q2$ and $\etjb$ are shown in figure~\ref{two}.

The measured cross section has a steep fall-off of five (four) orders of
magnitude over the measured range of $\q2$ ($\etjb$). NLO QCD
calculations using DISENT, the MRST99~\cite{mrs} set of proton PDFs and
different choices of $\mu_R$ have been compared to the data. The
calculations describe reasonably well the $\q2$ ($\etjb$) dependence
of the cross section for $\q2>500$ \g2\ ($\etjb>15$ GeV).

Inclusive jet cross sections have been measured in the laboratory frame
in NC \cite{jetdis} and CC \cite{cc} interactions to search for deviations
from the SM. The jets have been reconstructed using the
$\kt$ cluster algorithm in the kinematic region of $\q2>125\ (200)$ \g2
and at least one jet of $\etjet>14\ (8)$ GeV and $\etarr$ for NC
(CC) interactions is required.

\begin{figure}[h]
\setlength{\unitlength}{1.0cm}
\begin{picture} (18.0,10.0)
\put (2.0,-3.5){\epsfig{figure=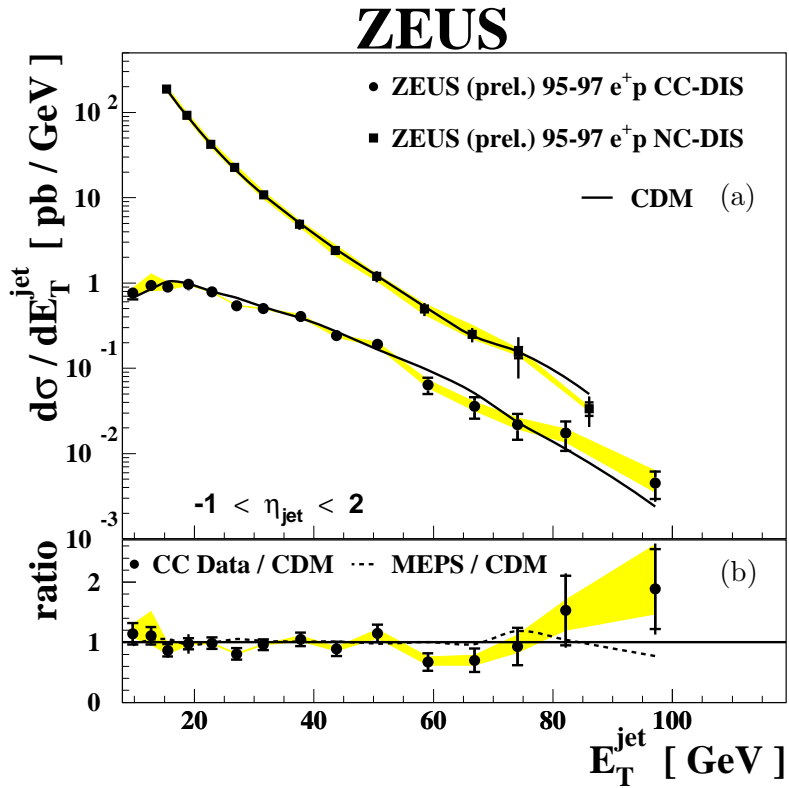,width=12cm}}
\put (11.7,7.5){\small (a)}
\put (11.7,2.5){\small (b)}
\end{picture}
\caption{\label{three}
{(a) Inclusive jet cross section in NC and CC DIS as a function of
$\etjet$. (b) Ratio data/theory as a function of $\etjet$ in CC DIS.}}
\end{figure}

The inclusive jet cross sections as a function of $\etjet$ for NC and CC
interactions are shown in figure~\ref{three}. The cross section in NC
interactions displays a steep fall-off of four orders of magnitude over
the measured range. The behaviour of the $\etjet$ distribution in CC
interactions is very different: it is approximately constant at low
$\etjet$ and falls less rapidly than in NC, and it approaches the NC cross
section for $\etjet\sim 80$ GeV. This behaviour can be interpreted as
due to the presence of a massive propagator in CC events, as it has
already been observed in measurements of the $\q2$ dependence of the
inclusive CC DIS cross section \cite{ccnc}. The cross sections for jets
give an independent measurement of the different electroweak boson
propagators in NC and CC DIS.

The predictions of Monte Carlo models using either the color-dipole model
(CDM) \cite{cdm} or first-order QCD matrix elements plus parton-showers
(MEPS) \cite{lepto} have been compared to the measurements. The
calculations describe reasonably well the measured cross sections. No
significant deviation from the SM is observed in NC or CC interactions up
to the highest $\etjet$ measured value ($\sim 100$ GeV).

\section{Dijet cross sections}

Differential cross sections for dijet production in NC DIS have been
measured \cite{dijet} in the Breit frame for $470<\q2<20000$ \g2. In
this kinematic region, the experimental uncertainties on the
reconstruction of both the positron and the hadronic final state are
smaller than at lower $\q2$. In addition, the theoretical
uncertainties due to the modelling of the hadronic final state, to the
proton PDFs and to the higher-order contributions are minimised. For
these measurements, the $\kt$ cluster algorithm was run in the
longitudinally invariant inclusive mode in the Breit frame. Two jets
of $E_{T,{\rm B}}^{\rm jet,M}>8$ GeV, $E_{T,{\rm B}}^{\rm jet,m}>5$ GeV and
$-1<\etajet_{\rm Lab}<2$ in each event were required, where
$E_{T,{\rm B}}^{\rm jet,M}\ (E_{T,{\rm B}}^{\rm jet,m})$ is the
transverse energy of the jet in the Breit frame with the highest
(second highest) transverse energy in the event. The use of asymmetric
cuts on $E_{T,{\rm B}}^{\rm jet}$ avoids infrared sensitive regions where
the behaviour of the cross section as predicted by the NLO QCD
programs is unphysical.

\begin{figure}[h]
\setlength{\unitlength}{1.0cm}
\begin{picture} (18.0,14.0)
\put (1.0,0.0){\epsfig{figure=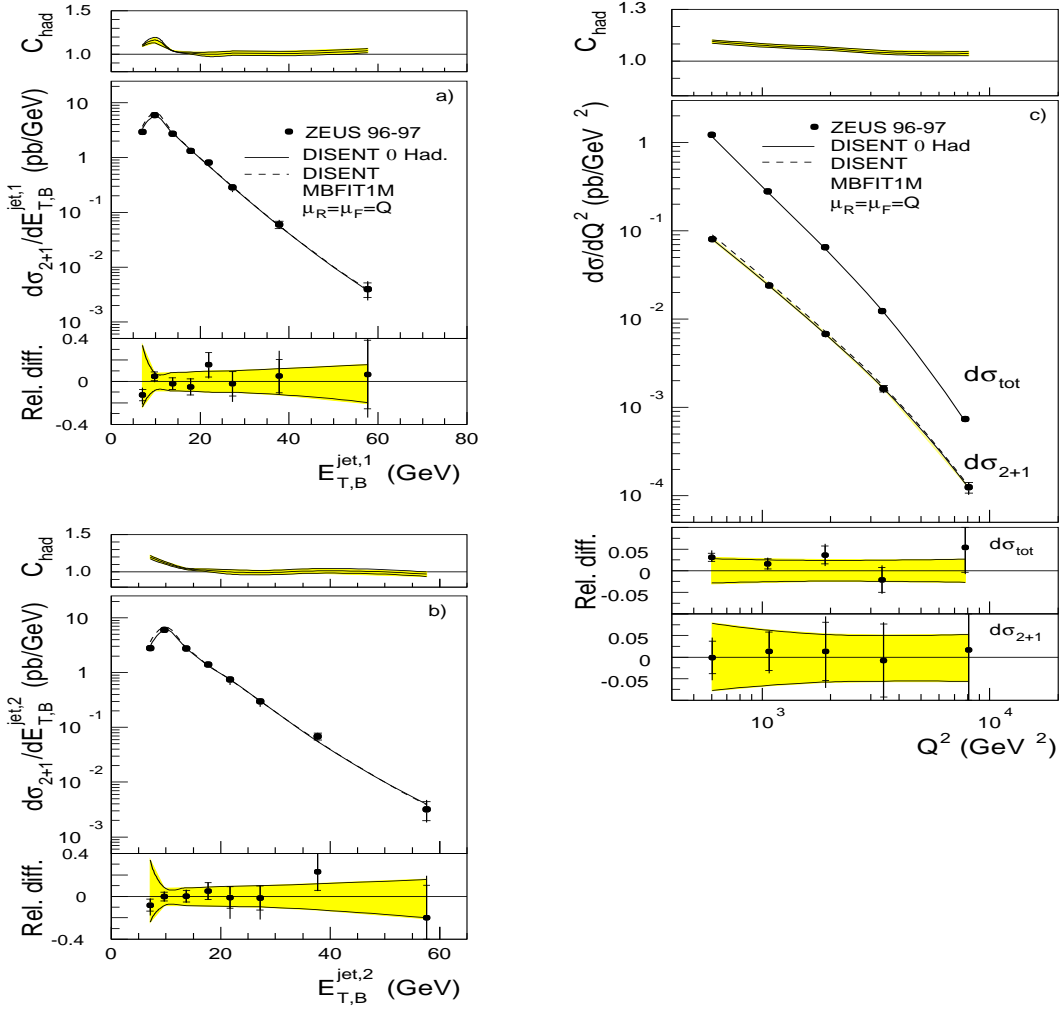,width=14cm,height=14cm}}
\end{picture}
\caption{\label{four}
{Dijet cross sections in NC DIS as functions of (a)
$E_{T,{\rm B}}^{\rm jet,1}$, (b)
$E_{T,{\rm B}}^{\rm jet,2}$ and (c) $\q2$. In (c), the inclusive differential
cross section ($d\sigma_{\rm tot}$) is also shown as a function of $\q2$.}}
\end{figure}

\begin{figure}[h]
\setlength{\unitlength}{1.0cm}
\begin{picture} (18.0,10.0)
\put (1.0,0.0){\epsfig{figure=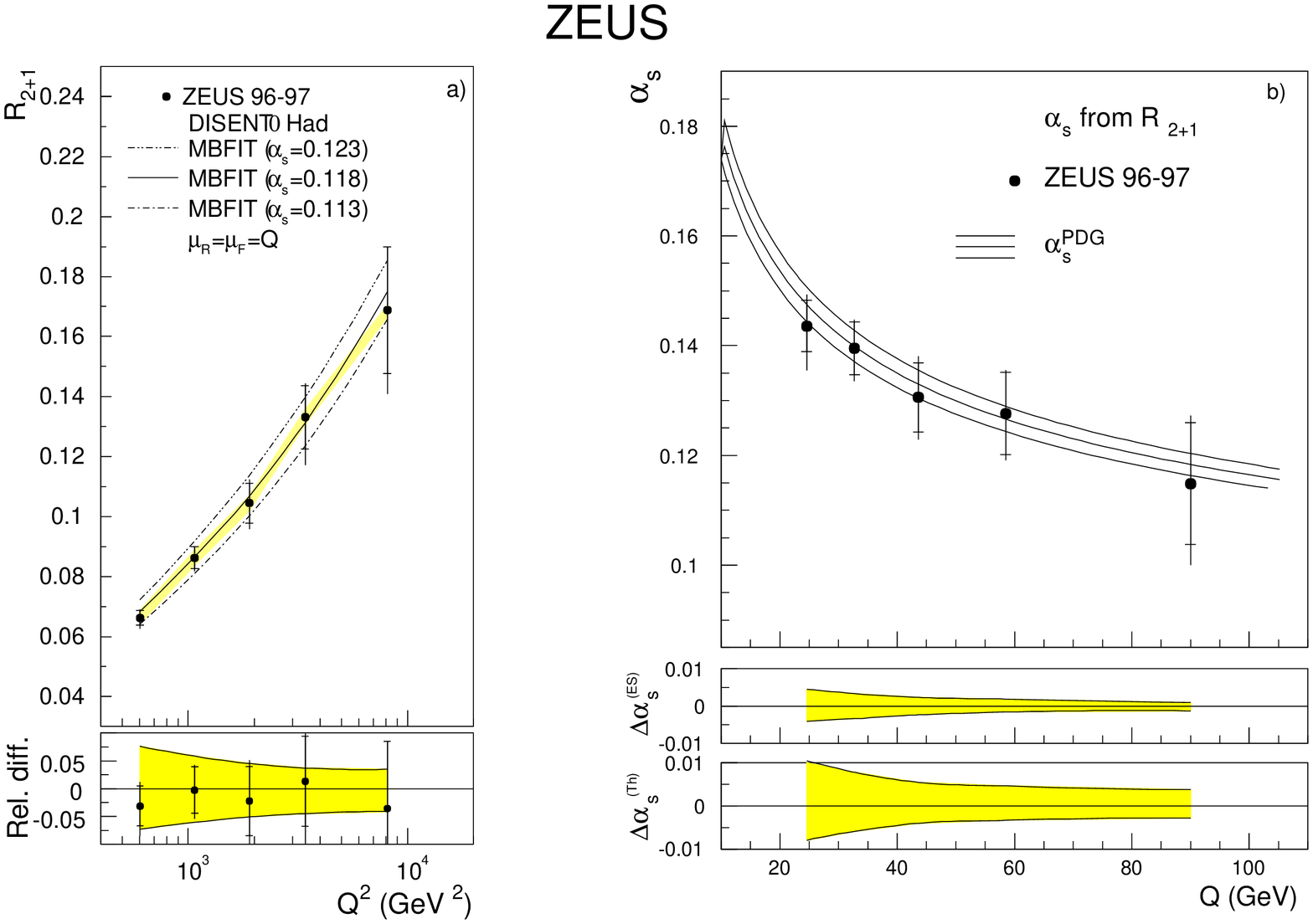,width=14cm}}
\end{picture}
\caption{\label{five}
{(a) The dijet fraction, $R_{2+1}$, in NC DIS as a function of $\q2$.
(b) The $\as$ values determined from the QCD fit of the measured dijet
fraction $R_{2+1}$ as a function of $Q$.}}
\end{figure}

The differential dijet cross sections as functions of
$E_{T,{\rm B}}^{\rm jet,(1,2)}$, where jet 1 (2) is the jet with highest
(lowest) pseudorapidity in the Breit frame, and $\q2$ are presented in
figure~\ref{four}. NLO QCD predictions calculated using DISENT are
compared to the measurements. The predictions, which assume
$\asz=0.118$, provide a good overall description of both the shape and
magnitude of the measured cross sections. The dijet fraction,
$R_{2+1}$, defined as the ratio of the dijet to the inclusive cross
section (see figure \ref{five}(a)), has been measured as a function of
$\q2$. The dijet fraction increases with increasing $\q2$ due to
phase-space effects. For the cross sections as a function of the jet
transverse energies and $\q2$, there is agreement at the
$\approx 10\%$ level between data and theory over four orders of
magnitude, demonstrating the validity of the  description of the
dynamics of dijet production by the NLO QCD hard processes.

\begin{figure}[h]
\setlength{\unitlength}{1.0cm}
\begin{picture} (18.0,8.0)
\put (-0.5,-3.0){\epsfig{figure=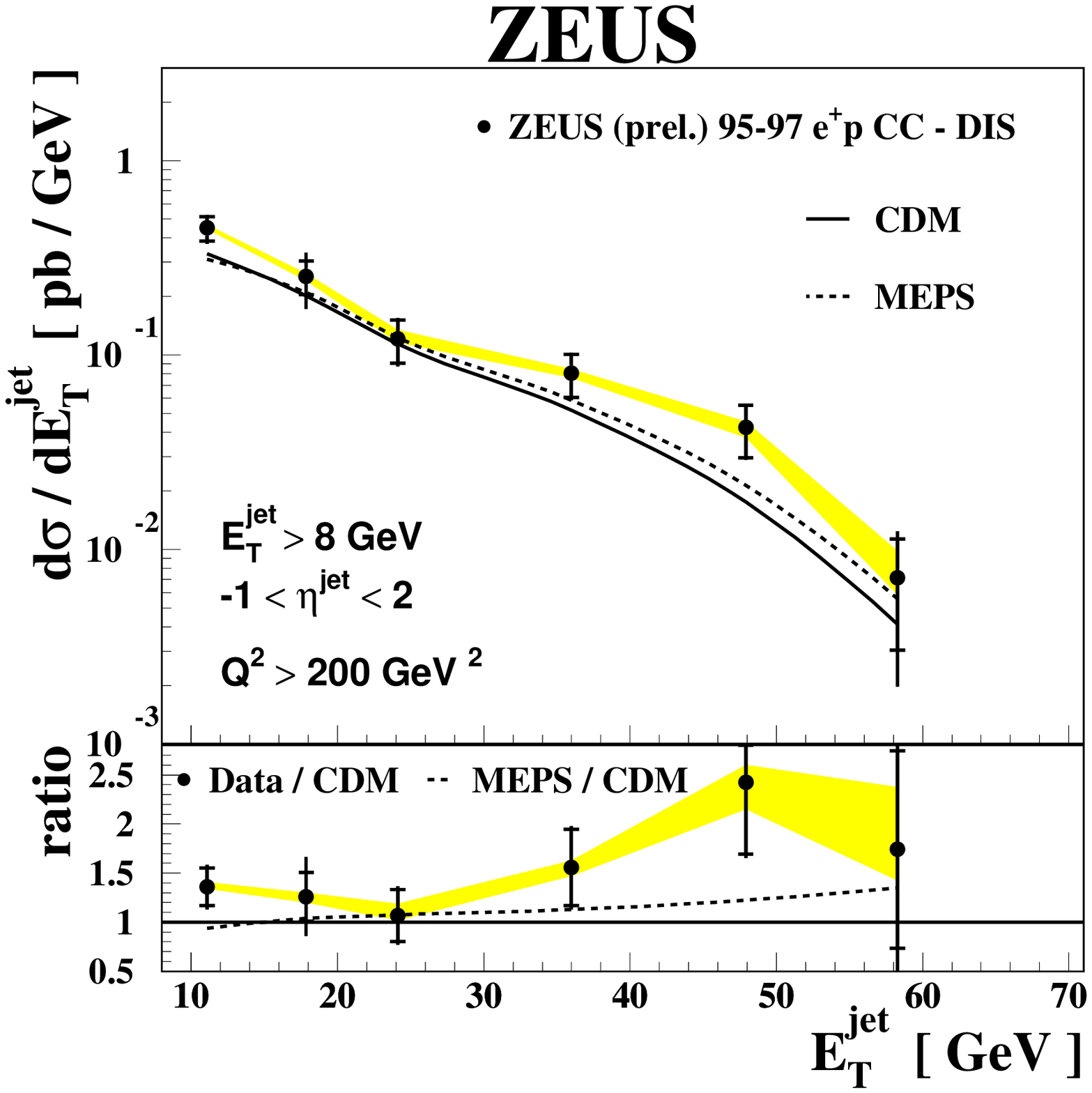,width=10cm}}
\put (8.3,-3.0){\epsfig{figure=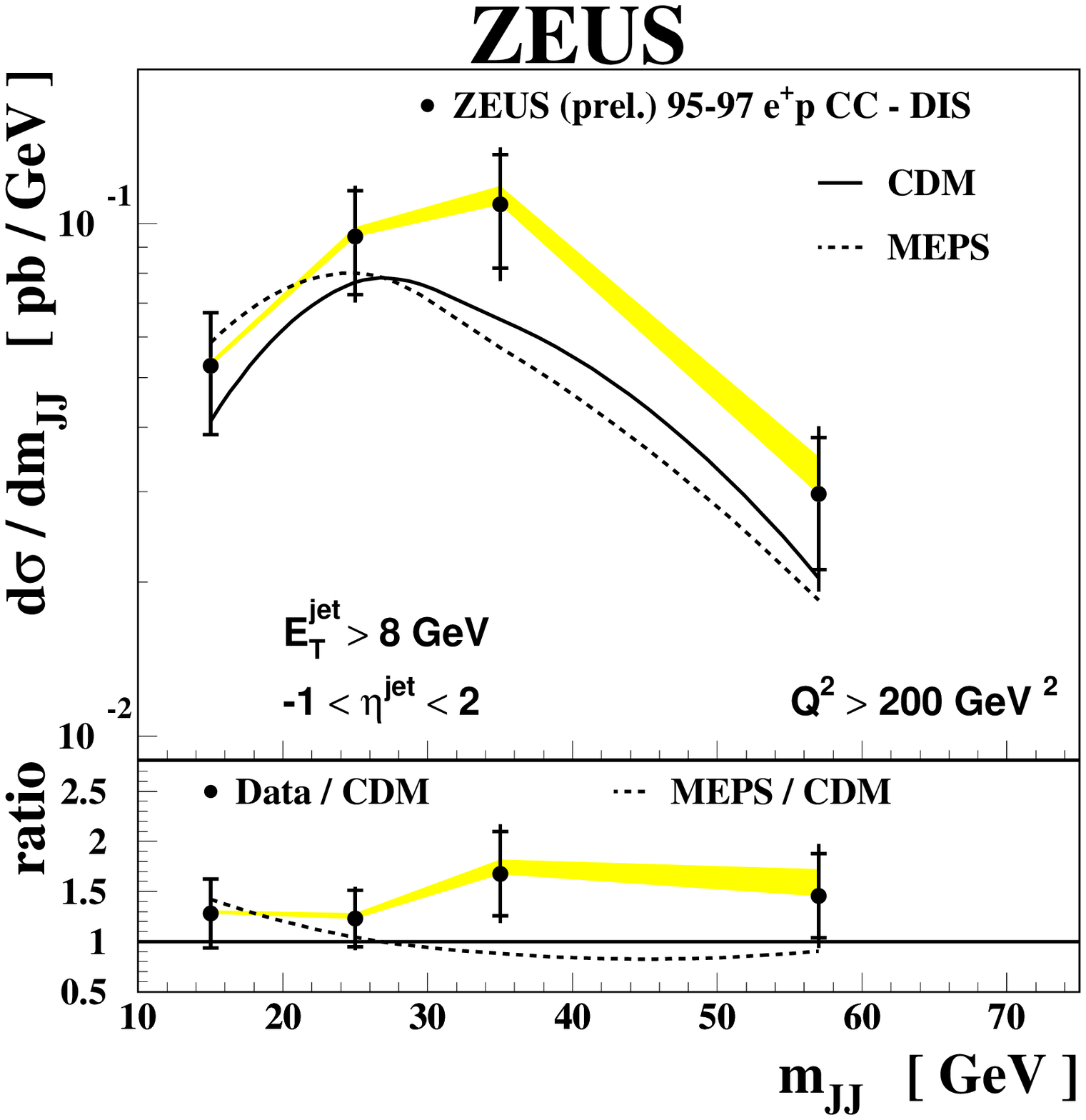,width=10cm}}
\put (7.5,6.1){\small (a)}
\put (7.5,2.0){\small (b)}
\put (16.3,6.1){\small (c)}
\put (16.3,2.0){\small (d)}
\end{picture}
\caption{\label{six}
{Dijet cross sections in CC DIS as a function of $\etjet$ (a) and
$\mjj$ (c). Ratio data/theory as a function of $\etjet$ (b) and $\mjj$
(d).}}
\end{figure}

Dijet cross sections in CC interactions have been measured \cite{cc} in the
laboratory frame as a function of the transverse energy of each of the
two highest $\etjet$ jets (figure~\ref{six}(a)) and as a function of the
invariant mass, $\mjj$, of these two jets (figure~\ref{six}(b)). The jets
have been reconstructed using the $\kt$ cluster algorithm in the kinematic
region of $\q2>200$ \g2 and at least two jets of $\etjet>8$ GeV and
$\etarr$ are required. Values as high as 60 GeV for $\etjet$ and
$\mjj$ are accessible with the luminosity used in this analysis. The
shapes of the $\etjet$ distributions at low $\etjet$ in inclusive jet (see
figure~\ref{three}) and dijet production are very different: this is
interpreted as the inclusive jet cross section being governed by
electroweak interactions and the heavy mass of the boson propagator at
low $\etjet$, whereas the dijet cross section is mainly dictated by
QCD. The predictions of the CDM and MEPS models describe reasonably
well the shape of the measured dijet cross sections, but they
underestimate the normalisation by $\sim 50\%$; this is understood
since these models only include up to $\oas$ diagrams.

\section{Jet substructure}

\begin{figure}[h]
\setlength{\unitlength}{1.0cm}
\begin{picture} (18.0,4.0)
\put (4.0,0.0){\epsfig{figure=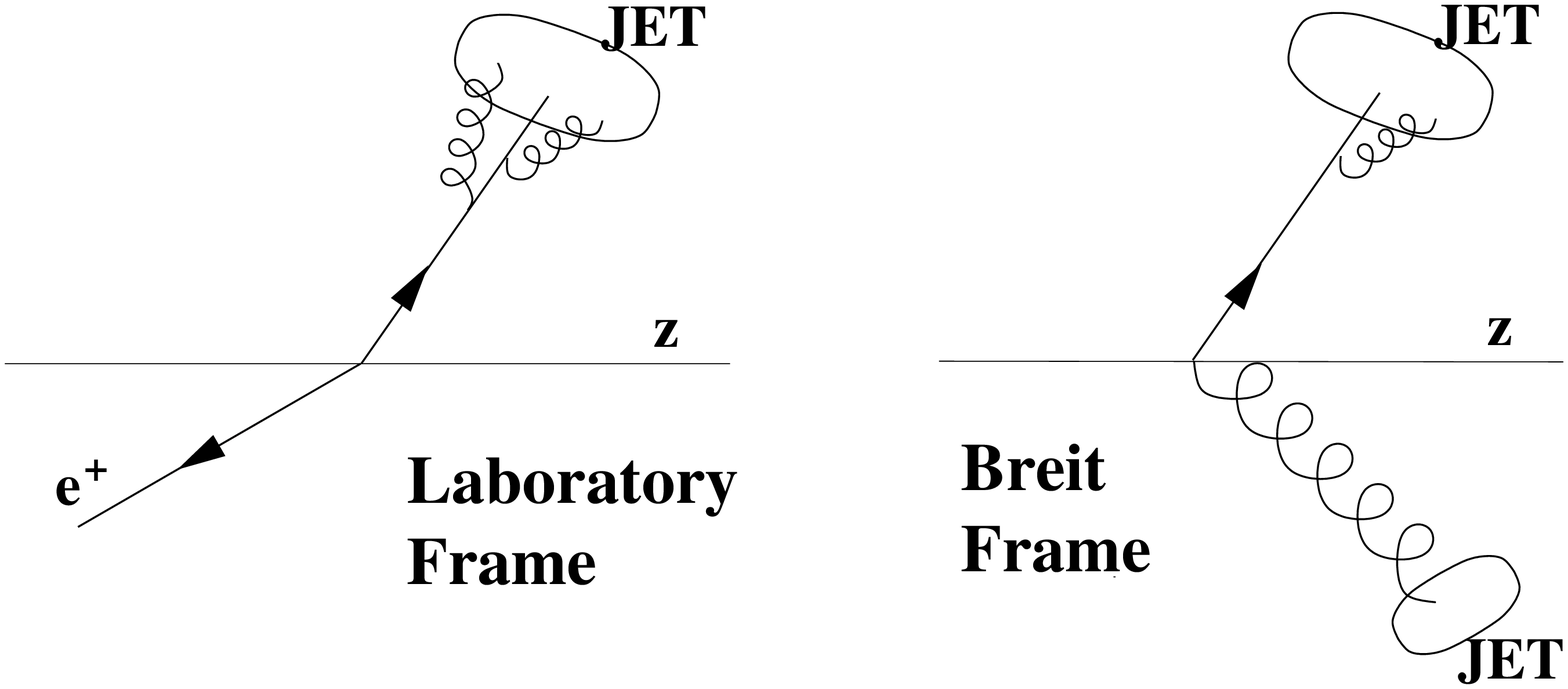,width=8cm}}
\end{picture}
\caption{\label{seven}
{Diagrams in $\oalphas2$ calculations.}}
\end{figure}

The investigation of the internal structure of jets gives insight into
the transition between a parton produced in a hard process and the
experimentally observed spray of hadrons. At sufficiently high
$\etjet$, where fragmentation effects become negligible, the jet
structure is expected to be calculable in pQCD. The lowest non-trivial
order contribution to the measurements of jet substructure is given by
$\oalphass$ calculations. Thus, measurements of jet substructure
provide a stringent test of pQCD and allow a determination of $\as$ by
comparing NLO calculations to the measurements. At present, this is
only possible for jets defined in the laboratory frame since only in
this frame can three partons be inside one jet (see figure~\ref{seven}).

\begin{figure}[h]
\setlength{\unitlength}{1.0cm}
\begin{picture} (18.0,12.5)
\put (1.0,-0.5){\epsfig{figure=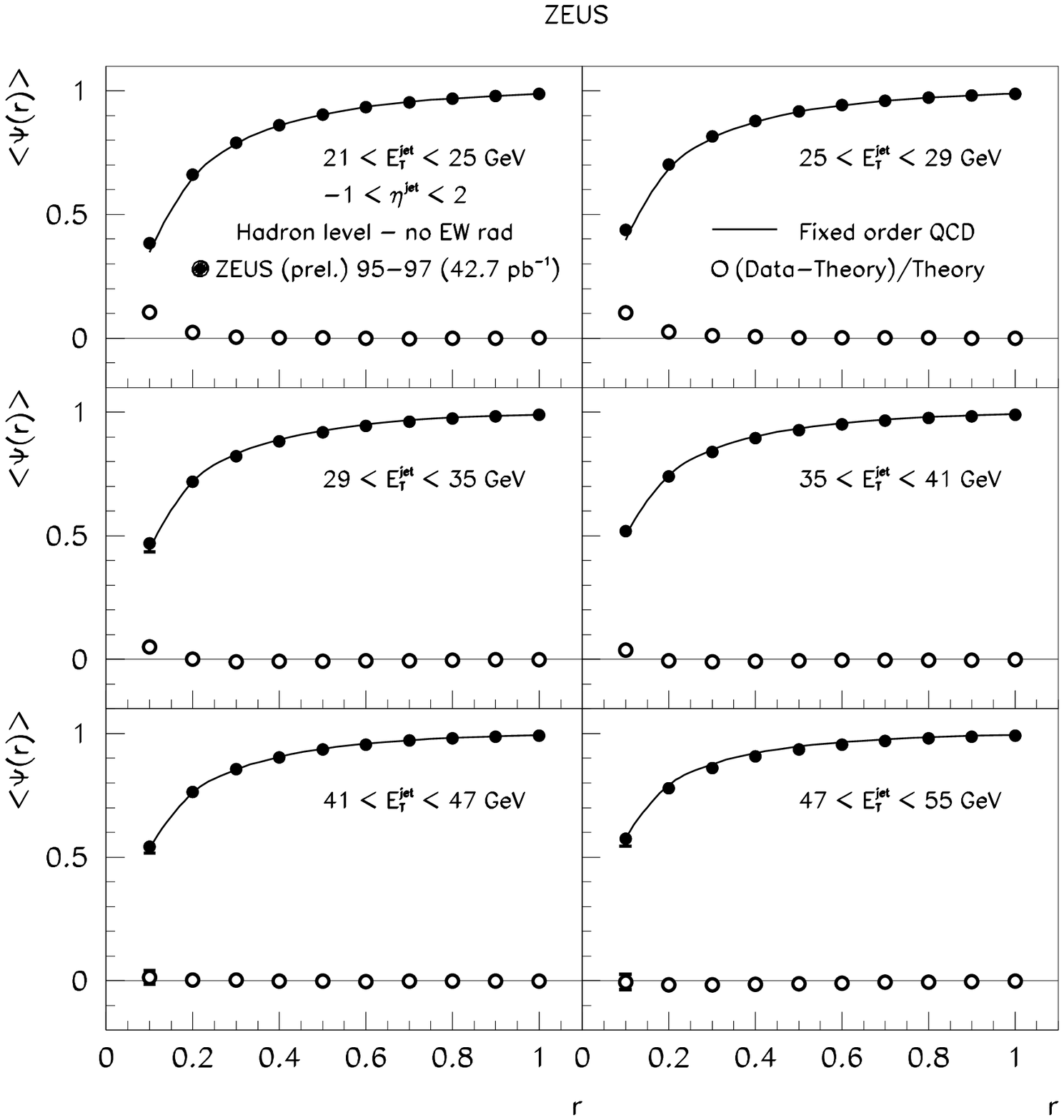,width=15cm}}
\end{picture}
\caption{\label{eight}
{Integrated jet shape as a function of $r$ in different regions of
$\etjet$ in NC DIS.}}
\end{figure}

The jet substructure was studied by means of the integrated jet shape
in inclusive jet production in NC interactions \cite{alphas} using the
$\kt$ cluster algorithm in the laboratory frame. The integrated jet
shape is defined as the average fraction of the jet's transverse
energy that lies inside a cone in the $\etaphi$ plane of radius $r$
concentric with the jet axis,
$$\langle \psi(r)\rangle \ =\ \frac{1}{N_{\rm jets}}\sum_{\rm jets}\frac{E_T(r)}{\etjet},$$
where $E_T(r)$ is the transverse energy associated with the jet within the
given cone of radius $r$ and $N_{\rm jets}$ is the total number of jets in the
sample. Only the particles assigned to the jet are considered. The
integrated jet shape as a function of the radius $r$ in different regions
of $\etjet$ is shown in figure~\ref{eight}. 

In pQCD, $\langle 1-\psi(r)\rangle$ is calculated as the fraction of the
jet's transverse energy, due to parton emission, that lies outside of
the cone of radius $r$,
$$\langle 1-\psi(r)\rangle ={\int dE_T\  E_T \ [d\sigma (ep\rightarrow 2\ {\rm partons})/dE_T]\over \etjet\ \sigma_{\rm jet}(\etjet)},$$
where $\sigma_{\rm jet}(\etjet)$ is the cross section for inclusive jet
production. NLO QCD predictions for the integrated jet shape are derived
from the above formula by computing the numerator to
${\cal O}(\alpha\as^2)$ and the denominator to ${\cal O}(\alpha\as)$.
Calculations using DISENT have been compared to the data. They give a very
good description of the measurements for $r\geq 0.2$ (see
figure~\ref{eight}).

The jet substructure was also studied by means of the mean
subjet multiplicity in NC \cite{alphas} and CC \cite{cc}
interactions. Subjets are jet-like objects within a jet which are
resolved by reapplying the $\kt$ cluster algorithm until for every
pair of particles the quantity
$$d_{ij}=\min(E_{T,i},E_{T,j})^2[(\eta_i-\eta_j)^2+(\varphi_i-\varphi_j)^2]$$
is above $\yc\cdot(\etjet)^2$, where $\yc$ is the resolution parameter.

\begin{figure}[h]
\setlength{\unitlength}{1.0cm}
\begin{picture} (18.0,10.0)
\put (2.0,-0.5){\epsfig{figure=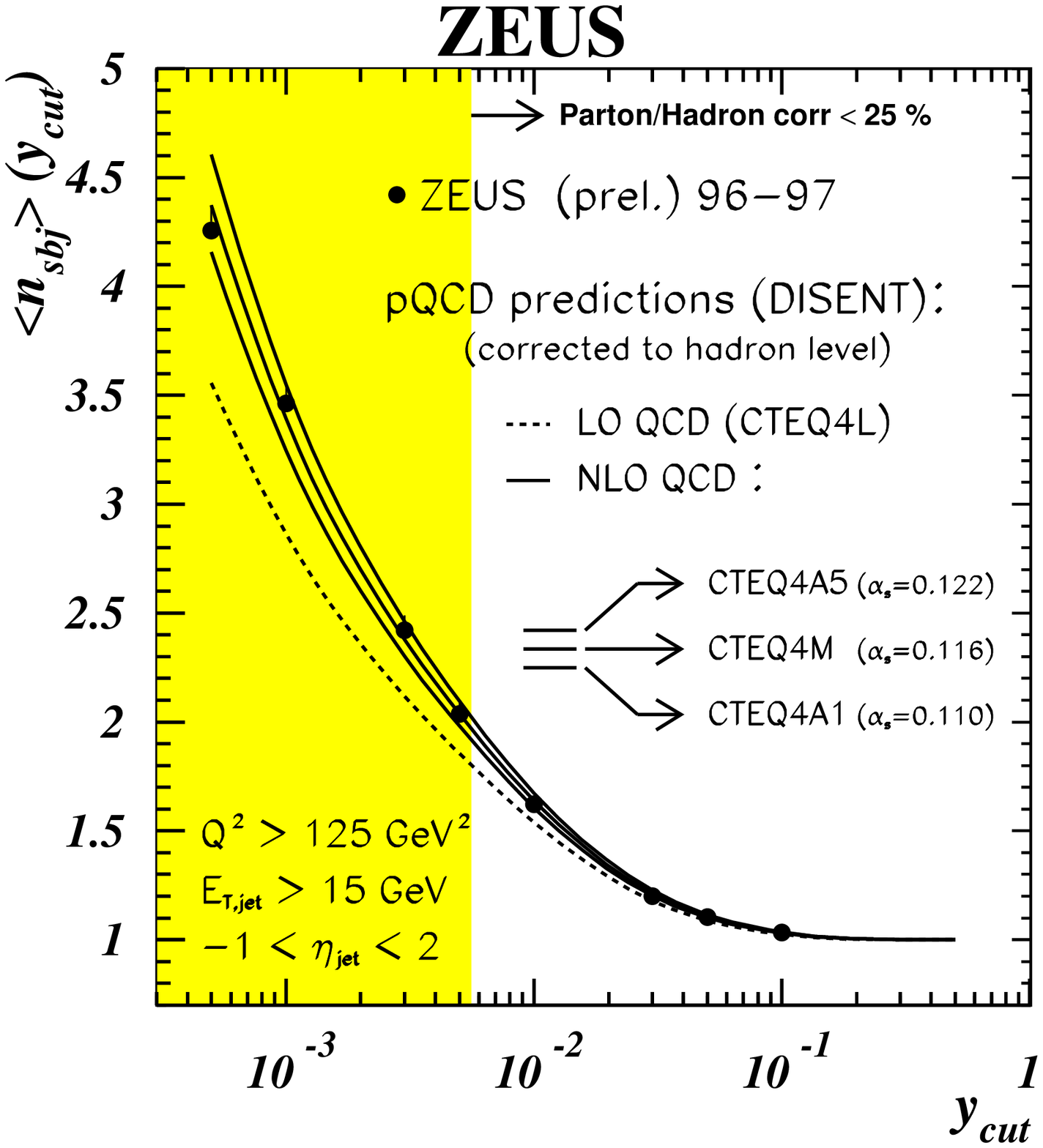,width=12cm}}
\end{picture}
\caption{\label{nine}
{Mean subjet multiplicity as a function of $\yc$ in NC DIS.}}
\end{figure}

The mean subjet multiplicity as a function of the resolution parameter
$\yc$ is presented in figure~\ref{nine} for NC interactions. In pQCD,
$\langle \ns-1\rangle$ is calculated as the ratio of the cross section
for $\ns-1$ subjets over that of inclusive jet production,
$$\langle \ns-1\rangle ={\sigma_{\rm \ns-1}(\etjet)\over \sigma_{\rm jet}(\etjet)}.$$
NLO QCD predictions for the mean subjet multiplicity are derived from the
above formula by adding 1 and computing the numerator (denominator) to
${\cal O}(\alpha\as^2)$ (${\cal O}(\alpha\as)$). NLO calculations computed
using DISENT and different parametrisations of the proton PDFs, which were
determined assuming different values of $\as$, have been compared to the
data. The NLO QCD calculation using the CTEQ4M proton PDFs and
$\asz=0.116$ gives a good description of the measurements.

\begin{figure}[h]
\setlength{\unitlength}{1.0cm}
\begin{picture} (18.0,10.0)
\put (3.0,-2.5){\epsfig{figure=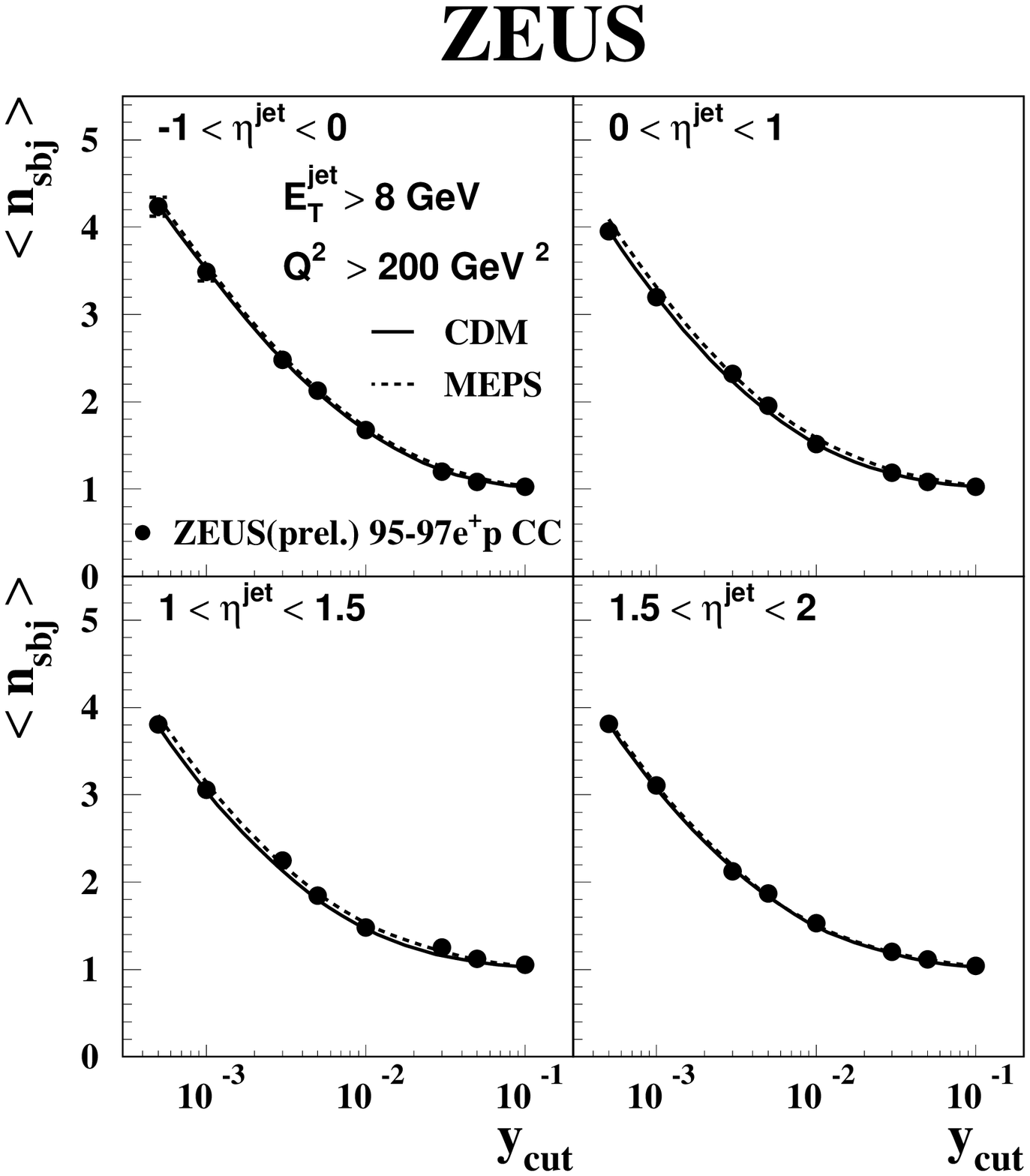,width=11.5cm}}
\end{picture}
\caption{\label{ten}
{Mean subjet multiplicity as a function of $\yc$ in different
$\etajet$ regions in CC DIS.}}
\end{figure}

\begin{figure}[h]
\setlength{\unitlength}{1.0cm}
\begin{picture} (18.0,10.0)
\put (3.0,-2.5){\epsfig{figure=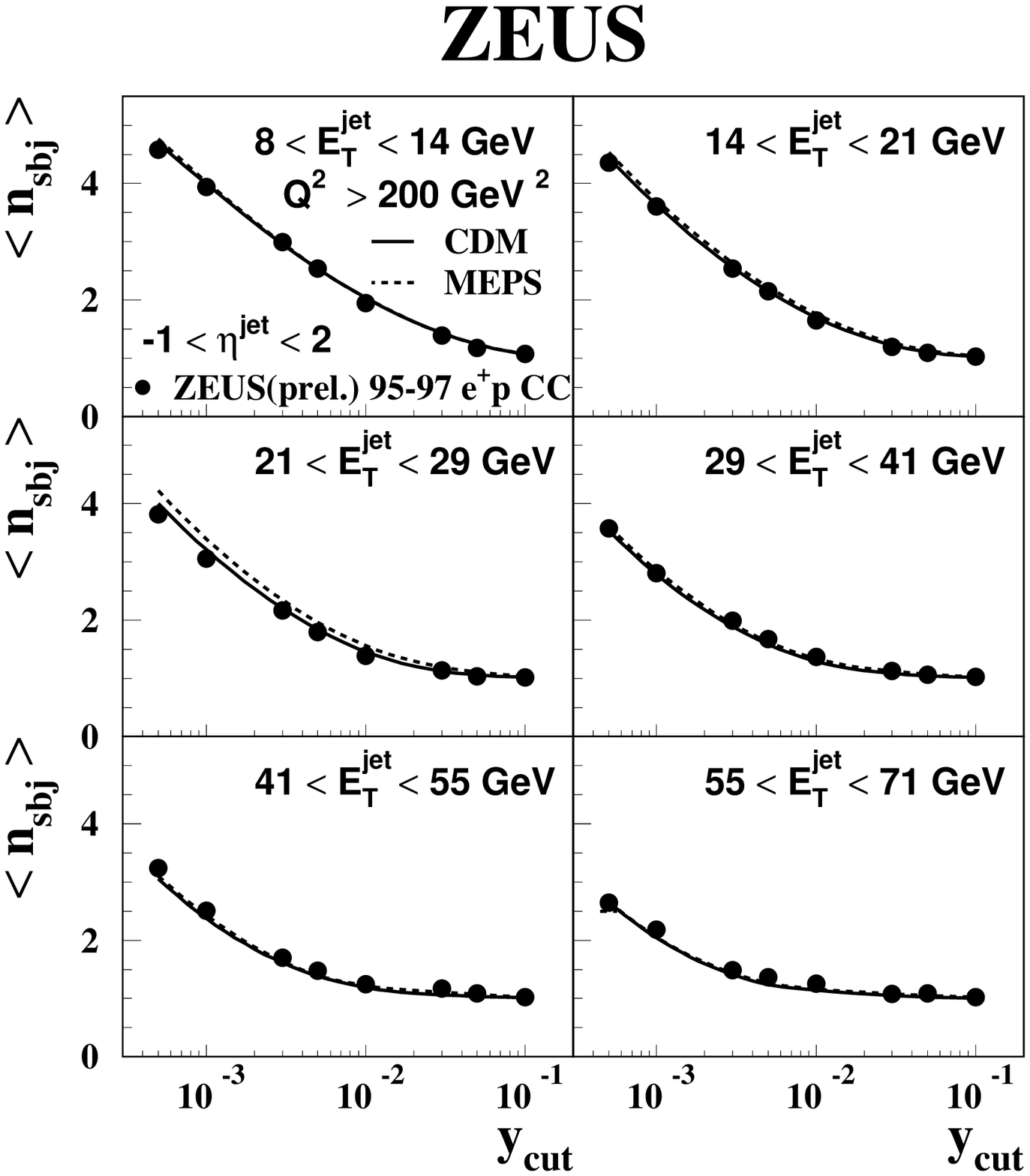,width=11.5cm}}
\end{picture}
\caption{\label{eleven}
{Mean subjet multiplicity as a function of $\yc$ in different $\etjet$
regions in CC DIS.}}
\end{figure}

The measured mean subjet multiplicity as a function of $\yc$ for
different regions in $\etajet$ and $\etjet$ is shown in figures~\ref{ten}
and \ref{eleven}, respectively, for CC interactions. The
predicted mean subjet multiplicities at the hadron level using the CDM
and MEPS models have been compared to the measurements. The measured
mean subjet multiplicities are well described by the predictions.

\begin{figure}[h]
\setlength{\unitlength}{1.0cm}
\begin{picture} (18.0,8.0)
\put (-2.0,-1.5){\epsfig{figure=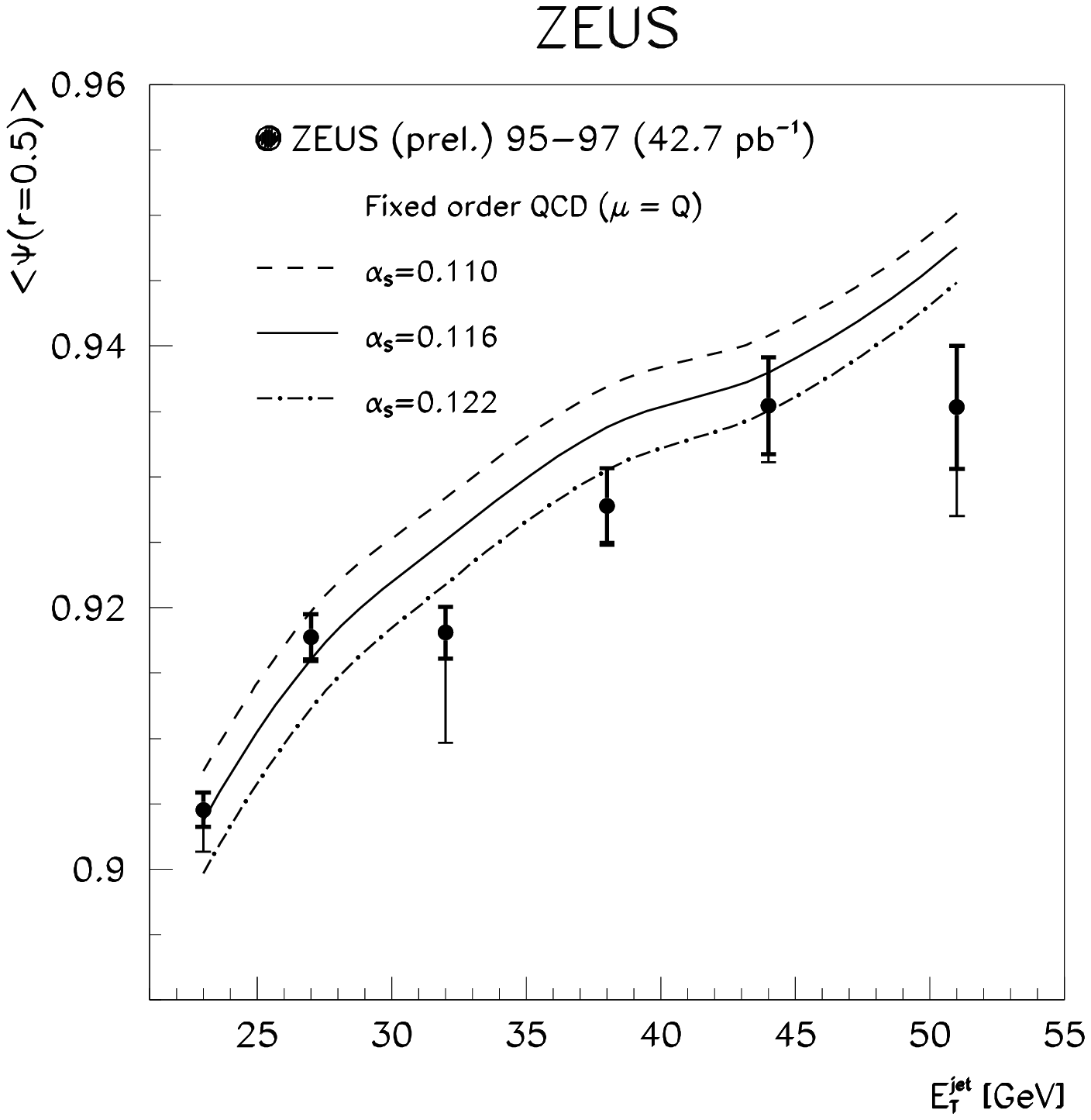,width=12.5cm}}
\put (8.0,-0.5){\epsfig{figure=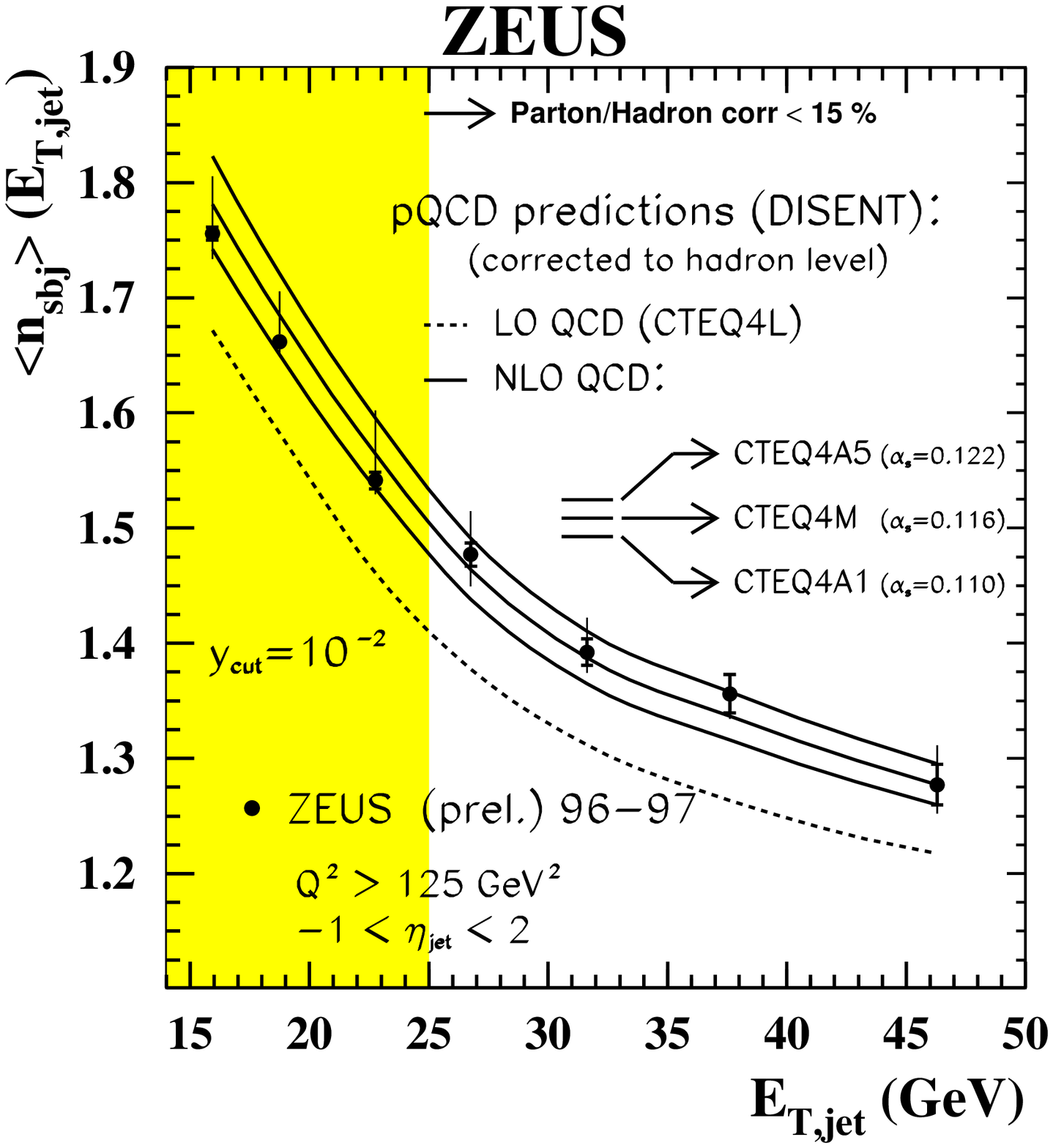,width=10cm}}
\put (7.0,8.0){\small (a)}
\put (16.2,8.0){\small (b)}
\end{picture}
\caption{\label{twelve}
{Integrated jet shape at a fixed value of $r=0.5$ (a) and mean subjet
multiplicity at a fixed value of $\yc=10^{-2}$ (b) as a function of
$\etjet$ in NC DIS.}}
\end{figure}

The $\etjet$ dependence of the integrated jet shape at a fixed value of
$r=0.5$ (figure~\ref{twelve}(a)) and that of the mean subjet
multiplicity at a fixed value of $\yc=10^{-2}$ (figure~\ref{twelve}(b))
in NC interactions have also been studied \cite{alphas}. The jet
shape increases and the subjet multiplicity decreases as $\etjet$
increases: the jets become narrower as $\etjet$ increases. NLO QCD
calculations using DISENT and different parametrisations of the PDFs
have been compared to the data. They give a good description of the
measurements and display the sensitivity of these observables to $\as$.

\begin{figure}[h]
\setlength{\unitlength}{1.0cm}
\begin{picture} (18.0,7.0)
\put (-1.0,-9.0){\epsfig{figure=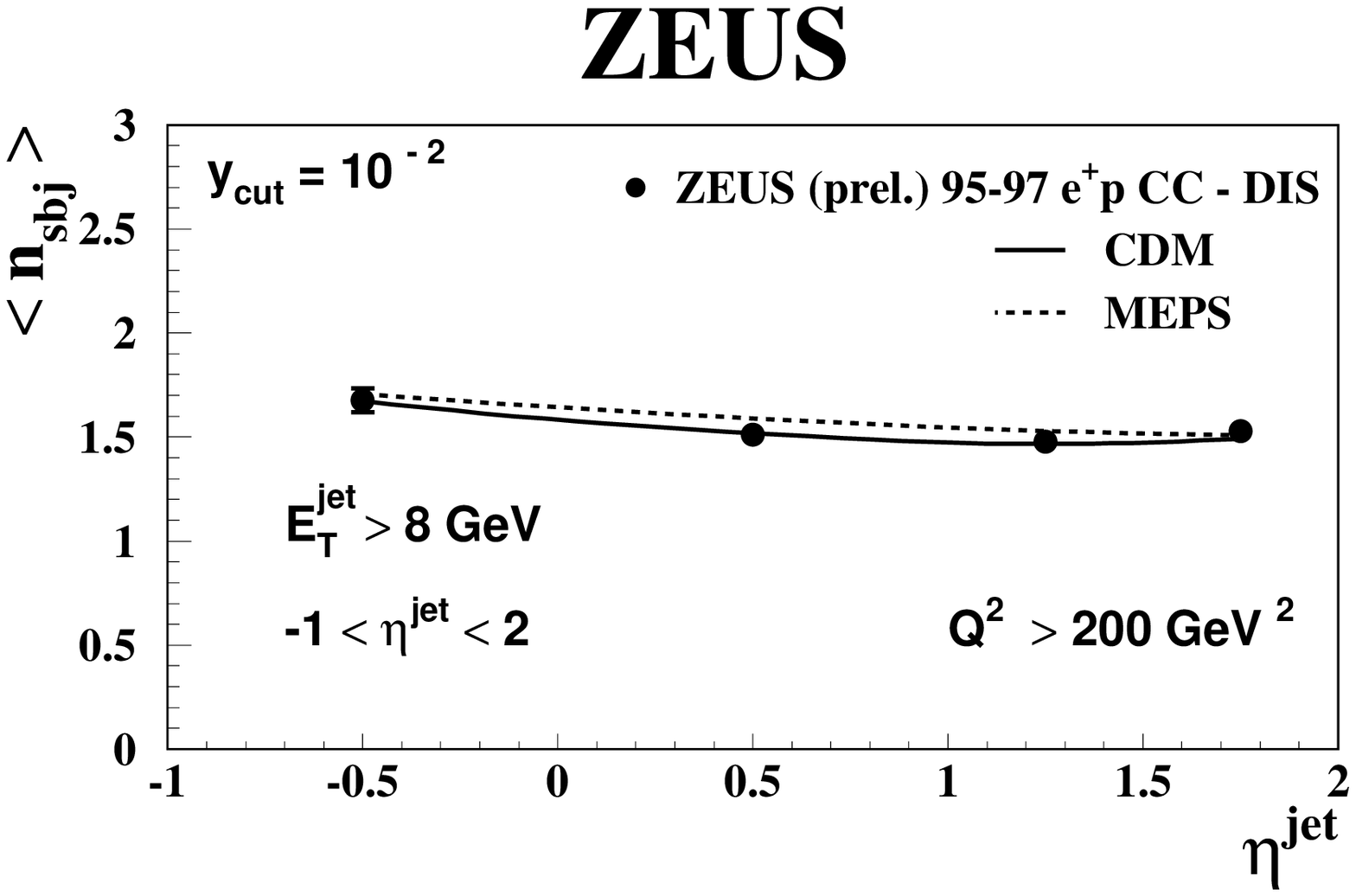,width=10cm,height=15cm}}
\put (8.5,-9.0){\epsfig{figure=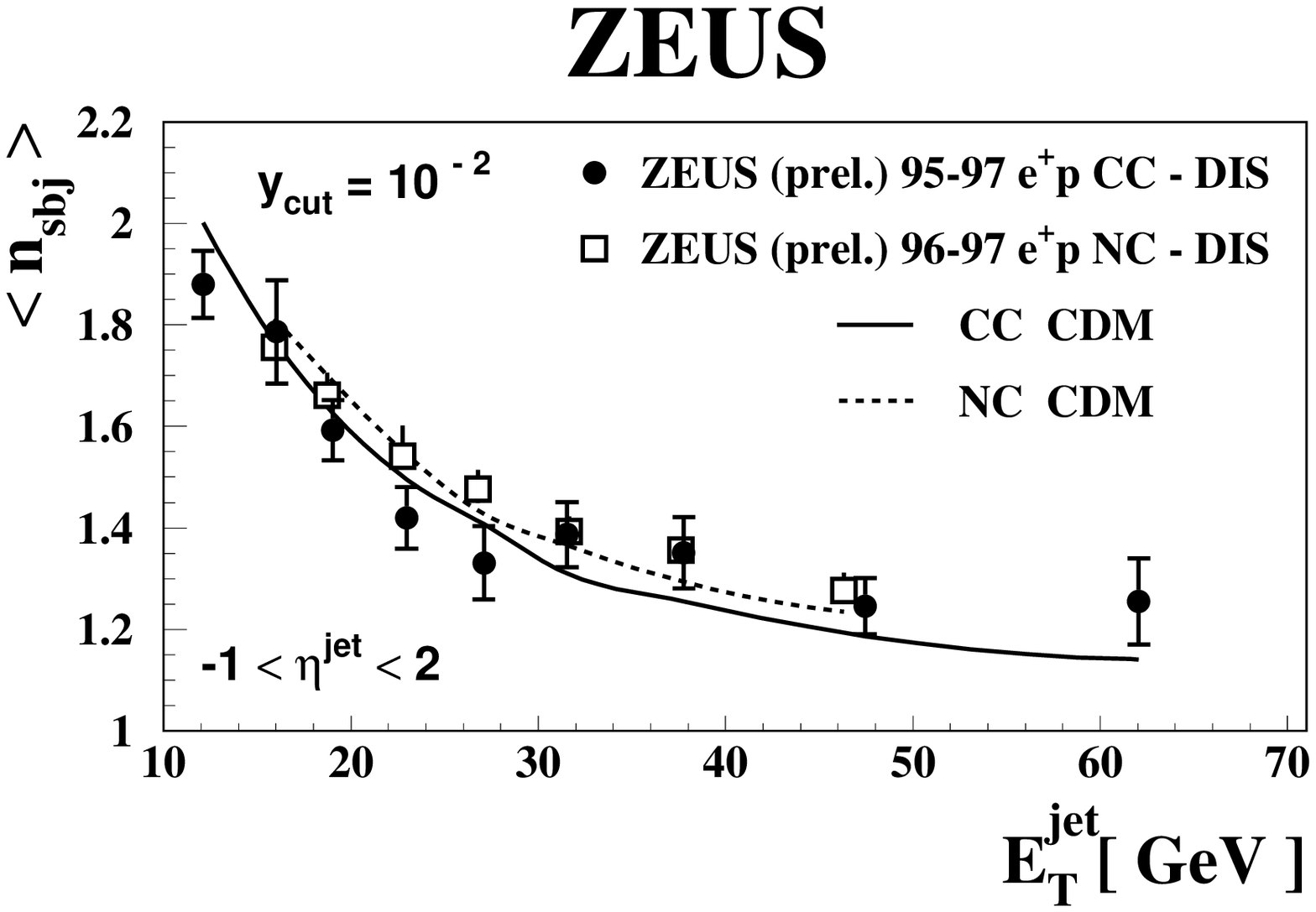,width=10cm,height=15cm}}
\put (7.0,1.8){\small (a)}
\put (16.5,1.8){\small (b)}
\end{picture}
\caption{\label{thirteen}
{Mean subjet multiplicity at a fixed value of $\yc=10^{-2}$ as a
function of (a) $\etajet$ and (b) $\etjet$ in CC DIS.}}
\end{figure}

Figure~\ref{thirteen}(a) shows the measured $\langle\ns\rangle$ at
$\yc=10^{-2}$ as a function of $\etajet$ in CC interactions. It
exhibits no significant dependence on $\etajet$, as in the case of the
jet shape \cite{shapedis}. The measured $\langle\ns\rangle$ at
$\yc=10^{-2}$ as a function of $\etjet$ is presented in
figure~\ref{thirteen}(b). It decreases as $\etjet$ increases and thus
the jets become narrower as $\etjet$ increases. In both figures, the
predictions from the CDM model are in good agreement with the data
whereas the predictions from the MEPS model are slightly above the
data. In figure~\ref{thirteen}(b), the measurements from
figure~\ref{twelve}(b) are also shown. The measured $\langle\ns\rangle$ at
$\yc=10^{-2}$ in CC and NC are very similar. This similarity can be
attributed to a large content of final-state quark jets in these two
processes and shows that the pattern of QCD radiation close to a
primary quark is to a large extent independent of the hard scattering
process.

\section{QCD analyses}

\subsection{Determination of $\as$}

In recent publications \cite{inclusive,dijet,alphas}, a new method to
determine $\as$ has been presented. The method consists of performing NLO
calculations of an observable using the five CTEQ4 sets of the
``A-series'' \cite{cteq}; the value of $\asz$ in each calculation is
that of the corresponding set of PDFs. The calculations for each
$\asz$ are used to parametrise the $\as$ dependence of the
observables, and then, from the measurements, a value of $\asz$ is
directly extrapolated. The measurements of the inclusive jet cross
section \cite{inclusive} and the dijet fraction \cite{dijet} as a
function of $\q2$ and the integrated jet shape \cite{alphas} and mean 
subjet multiplicity \cite{alphas} as a function of $\etjet$ have been
used to determine a value of $\asz$. 

\begin{figure}[h]
\setlength{\unitlength}{1.0cm}
\begin{picture} (18.0,6.0)
\put (-1.0,-0.5){\epsfig{figure=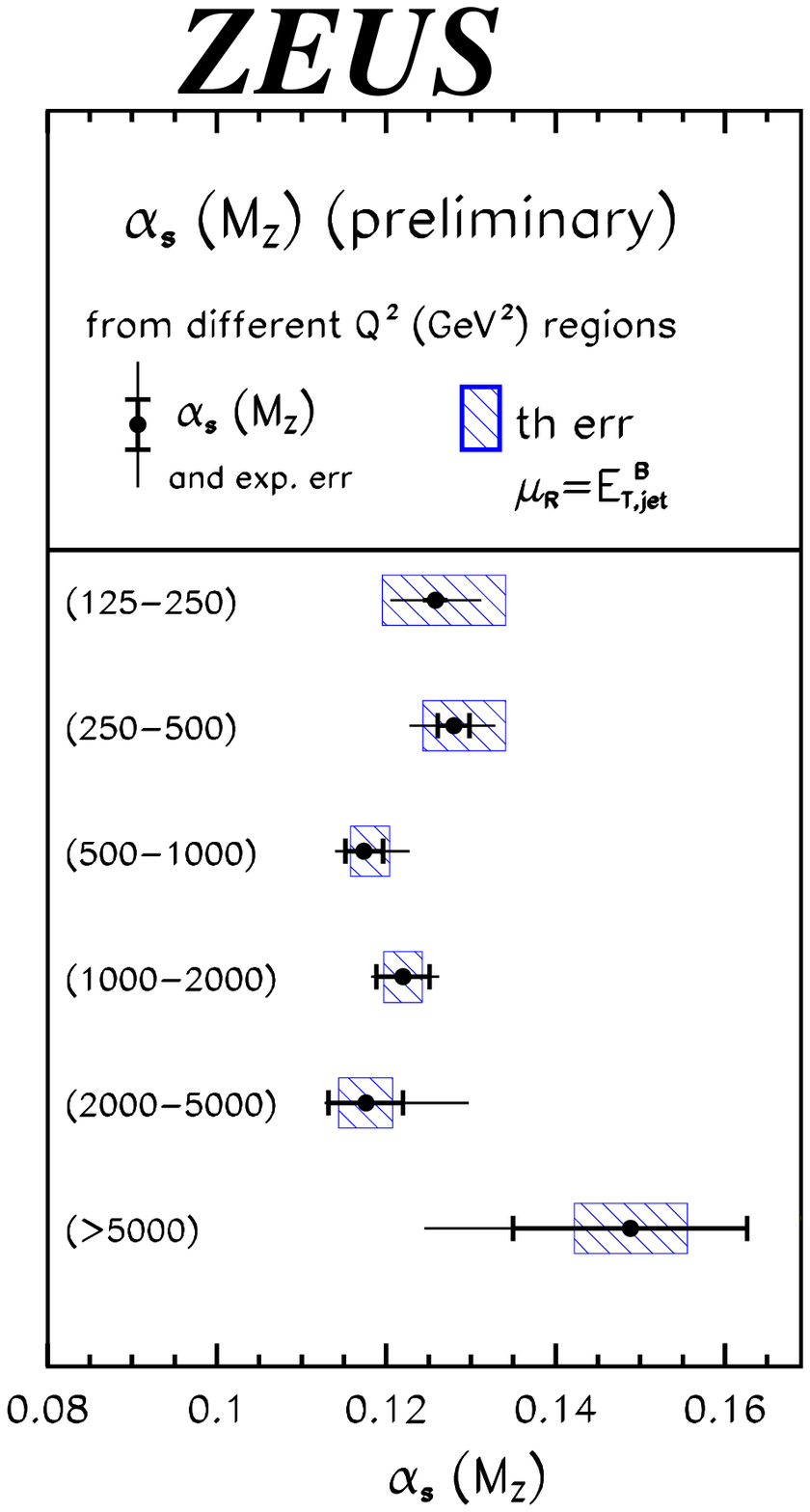,height=7cm,width=8cm}}
\put (4.0,-0.5){\epsfig{figure=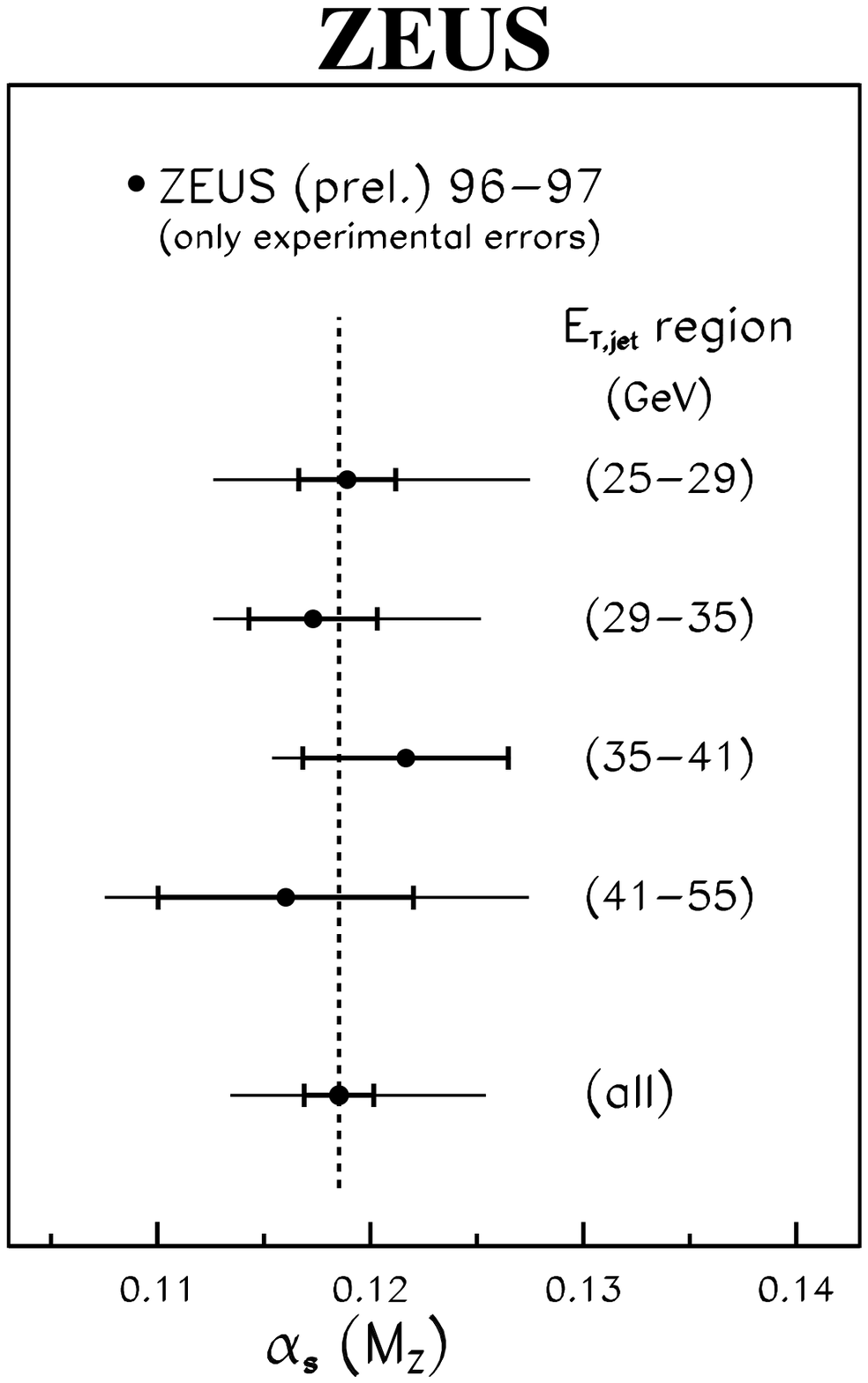,width=5cm,height=7cm}}
\put (8.0,-2.5){\epsfig{figure=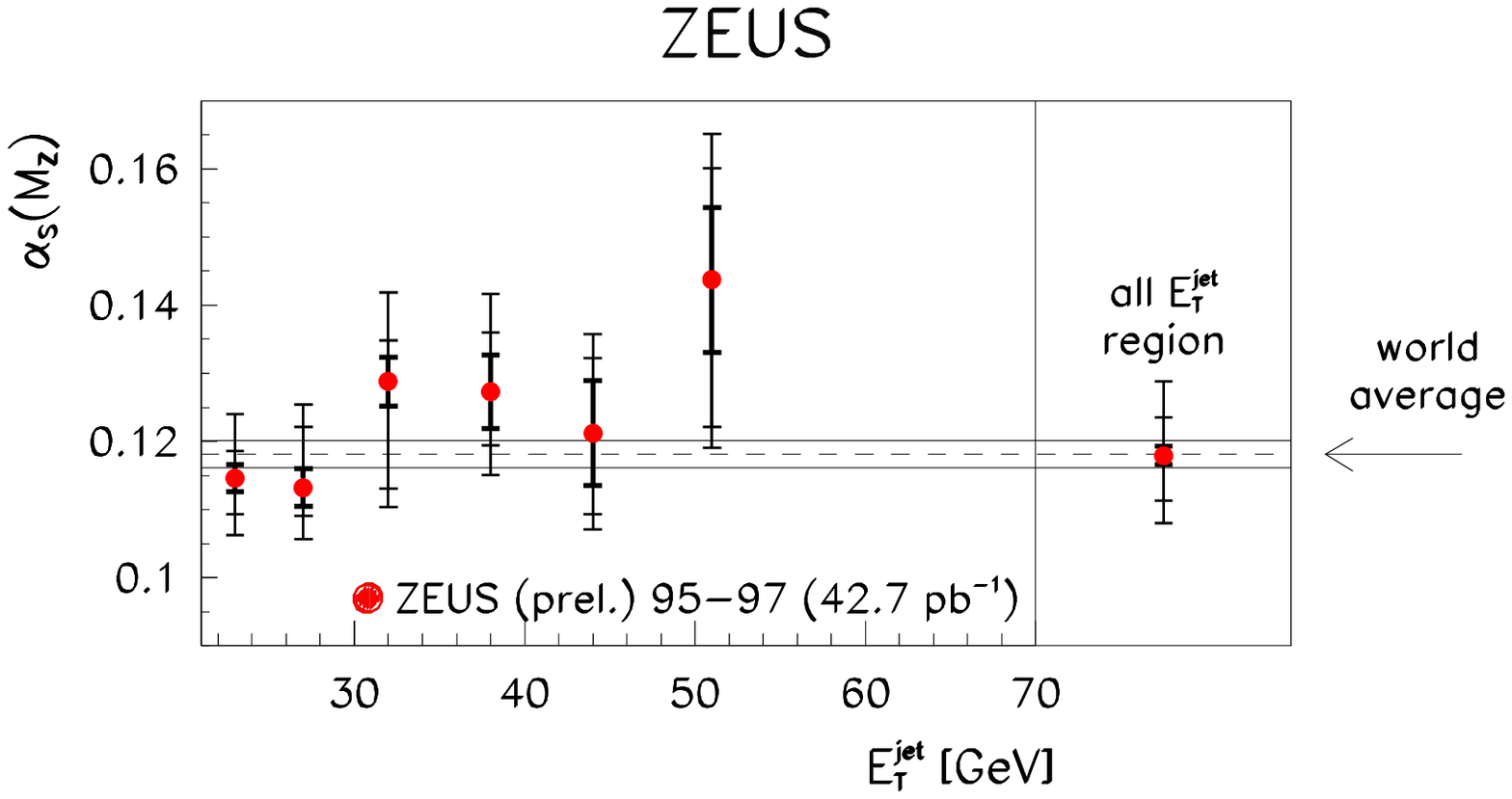,width=11cm}}
\put (3.6,5.3){\small (a)}
\put (7.5,5.3){\small (b)}
\put (16.0,4.0){\small (c)}
\end{picture}
\caption{\label{fourteen}
{Extracted values of $\asz$ from the inclusive jet cross section (a),
the mean subjet multiplicity (b) and the integrated jet shape (c) in
NC DIS.}}
\end{figure}

From each measurement in a given $\q2$ or $\etjet$ region, a value of
$\asz$ was determined. The results are shown in figure~\ref{fourteen}
separately for the inclusive jet cross section, the mean subjet
multiplicity and the integrated jet shape.

A combined value of $\asz$ from each observable has been obtained by
performing a $\chi^2$ fit to the data in the following kinematic ranges:
$\q2>500$ \g2\ for the inclusive jet cross section, $\q2>470$ \g2\ for
the dijet fraction, $\etjet>21$ GeV for the integrated jet shape and
$\etjet>25$ GeV for the mean subjet multiplicity. These are the
kinematic regions where the hadronisation correction factors to the
NLO calculations are small for each observable. The combined values of
$\asz$ determined from each observable are:
\vspace{0.25cm}

inclusive jet cross section:\hfill
$\asmz{0.1190}{0.0017}{0.0049}{0.0023}{0.0026}{0.0026}$;

\vspace{0.2cm}

dijet fraction:\hfill
$\asmz{0.1166}{0.0019}{0.0024}{0.0033}{0.0057}{0.0044}$;

\vspace{0.2cm}
integrated jet shape:\hfill
$\asmz{0.1179}{0.0014}{0.0054}{0.0065}{0.0094}{0.0073}$;

\vspace{0.2cm}
mean subjet multiplicity:\hfill
$\asmz{0.1185}{0.0016}{0.0067}{0.0048}{0.0089}{0.0071}$.

\vspace{0.5cm}

\begin{figure}[h]
\setlength{\unitlength}{1.0cm}
\begin{picture} (18.0,13.0)
\put (2.5,-0.5){\epsfig{figure=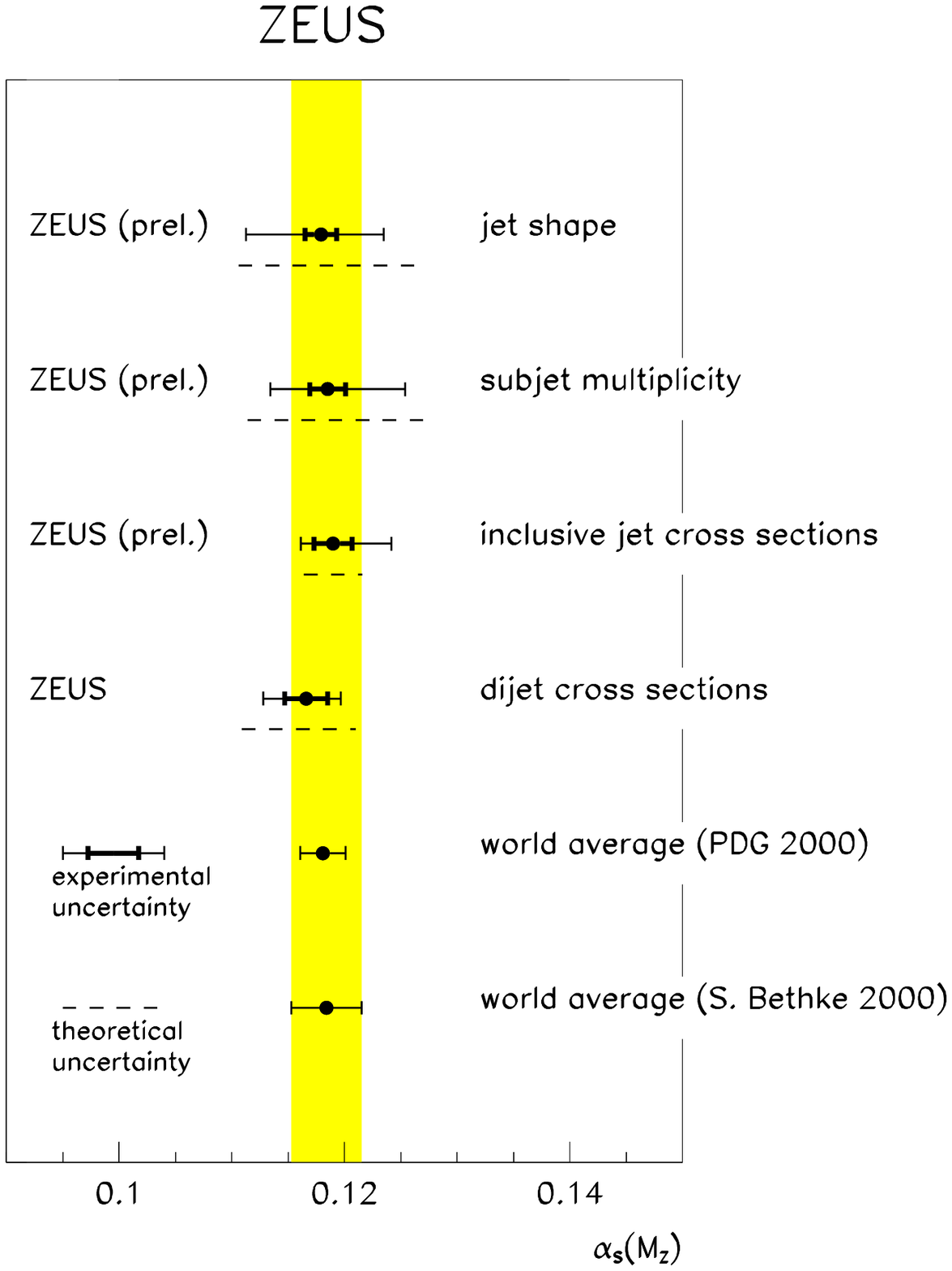,width=15cm}}
\end{picture}
\caption{\label{fifteen}
{A compilation of recent determinations of $\asz$ using jet production
in NC DIS.}}
\end{figure}

The values of $\asz$ obtained from the different observables are
compatible with each other and with the world average 
\cite{alpworld,bethke} (see figure~\ref{fifteen}).

\subsection{Energy scale dependence of $\as$}
The QCD fit of the dijet fraction has been repeated in five $\q2$ bins
to test the scale dependence of the renormalised strong coupling
constant. The method of the fit is the same as outlined above, but the
$\as$ dependence of the dijet fraction was parametrised in terms of
$\as(\langle Q \rangle)$, where $\langle Q \rangle$ is the mean value
of $Q$ in each bin. The measured $\as(\langle Q \rangle)$ values are
shown in figure~\ref{five}(b). The measurements are compared with the
renormalisation group predictions obtained from the PDG $\asz$ value
and its associated uncertainty. The values are in good agreement with
the predicted running of the strong coupling constant over a large
range in $Q$.

\section{Summary and conclusions}

Recent measurements of jet cross sections in neutral-current and
charged-current deep inelastic $ep$ scattering have been presented. From
QCD analyses of these measurements, determinations of the strong coupling
constant $\as$ have been obtained. The results are consistent with the
world average value of $\asz$. The measurements have been performed in
regions of phase space where the experimental uncertainties are small
and thus the determinations are precise.

However, the next-to-leading order (NLO) calculations used in the QCD
analyses have relatively large uncertainties. For more accurate
measurements of the parameters of QCD, improved calculations are
needed. For instance, higher-order corrections are required to reduce
the largest theoretical uncertainty, namely that of the terms beyond NLO.

\vspace{0.5cm}
\noindent {\bf Acknowledgments}.
This report is partially based on a talk given at the {\it International
Europhysics Conference on High Energy Physics}, Budapest, Hungary,
July 2001.
I would like to thank J. Terr\'on for a critical reading of the
manuscript.


\end{document}